\documentstyle[epsfig,12pt,preprint,tighten,aps]{revtex}
\begin{document}

\draft

\title{\rightline{{\tt March 1998}}
\rightline{{\tt UM-P-98/15}}
\rightline{{\tt RCHEP-98/04}}
\ \\
Energy-dependent solar neutrino flux depletion in the\\
Exact Parity Model and implications for \\
SNO, SuperKamiokande and BOREXINO}
\author{Raymond R. Volkas and Yvonne Y. Y. Wong}
\address{School of Physics\\
Research Centre for High Energy Physics\\
The University of Melbourne\\
Parkville 3052 Australia\\
(r.volkas@physics.unimelb.edu.au, ywong@physics.unimelb.edu.au)}
\maketitle

\begin{abstract} 

Energy-dependent solar neutrino flux reduction caused by the
Mikheyev--Smirnov--Wolfenstein (MSW) effect is applied to the 
Exact Parity Model. 
Several scenarios are possible, depending on the region of 
parameter space chosen.
The interplay between intergenerational MSW transitions and 
vacuum ``intragenerational''
ordinary--mirror neutrino oscillations is discussed. Expectations 
for the ratio of
charged to neutral current event rates at the Sudbury Neutrino 
Observatory (SNO) are
estimated. The implications of the various scenarios for the Boron 
neutrino energy
spectrum and BOREXINO are briefly discussed. The consequences of 
MSW-induced solar
neutrino depletion within the Exact Parity Model differ in 
interesting ways from the
standard $\nu_e \leftrightarrow \nu_{\mu, \, \tau}$ and 
$\nu_e \leftrightarrow \nu_s$ cases. 
The physical causes of
these differences are determined.

\end{abstract}

\section{Introduction}

In the Exact Parity Model (EPM) \cite{epm}, 
parity is an exact symmetry of nature
despite the $V-A$ character of weak interactions. Exact parity symmetry is
achieved by introducing parity or ``mirror'' partners for each of the
standard model fermions, Higgs bosons and gauge bosons. In general, colour
singlet and electromagnetically neutral particles in the standard sector
mix with their corresponding mirror states, leading to possibly
observable experimental effects. 

One of the most interesting possibilities in this regard is mixing between
ordinary
and mirror neutrinos \cite{fv95}. In part, the EPM is an explicit theory 
featuring
three effectively sterile light neutrino flavours in addition to $\nu_e$,
$\nu_{\mu}$ and $\nu_{\tau}$. We shall denote the mirror neutrino flavours
by $\nu'_e$, $\nu'_{\mu}$ and $\nu'_{\tau}$, where $\nu'_{\alpha}$ is the
parity partner of $\nu_{\alpha}$ $(\alpha = e, \mu, \tau)$. Exact parity
invariance imposes a simple and nontrivial constraint on standard--mirror
neutrino mixing: in the absence of intergenerational mixing, the mass
eigenstate neutrinos must be maximal mixtures of ordinary and mirror
neutrinos. This follows immediately from the requirement that parity
eigenstates must also be eigenstates of the Hamiltonian when parity is an
exact symmetry. The mass/parity eigenstates are given by
\begin{equation}
\label{epmmixing1}
\nu_{\alpha \pm} \equiv \frac{\nu_{\alpha L} \pm 
(\nu'_{\alpha R})^c}{\sqrt{2}},
\end{equation}
where $\nu_{\alpha \pm} \to \pm(\nu_{\alpha\pm})^c$ under a parity
transformation.
When intergenerational mixing is nonzero and $CP$ violation absent, the mass
eigenstates are simply linear combinations of the $\nu_{\alpha +}$ and,
separately, the $\nu_{\alpha -}$:
\begin{eqnarray}
\label{epmmixing2}
\nu_{i+} & = & \sum_{\alpha} U^{+}_{i\alpha} \nu_{\alpha +}, \nonumber\\
\nu_{i-} & = & \sum_{\alpha} U^{-}_{i\alpha} \nu_{\alpha -},
\end{eqnarray}
where $i = 1, \, 2, \, 3$ and $U^{\pm}_{i\alpha}$ are unitary mixing
matrices.
Exact parity symmetry forbids mixing between positive and negative parity
neutrinos in the vacuum.\footnote{If the {\it minimal} standard model is
extended by adding a mirror sector, then both neutrinos and mirror
neutrinos are massless and unmixed. We do not consider this case because it
is of little interest for neutrino phenomenology.}

The neutrino sector of the EPM is of great interest because it can explain
both the solar and atmospheric neutrino anomalies \cite{fv95}. The clearest 
case is
provided by the atmospheric neutrino anomaly. (Note that we will consider
the case of small intergenerational mixing in this paper, taking our cue
from the almost diagonal Kobayashi--Maskawa matrix in the quark sector.) 
The observed anomalous value
of the ratio $R$ of $\mu$-like to $e$-like events strongly suggests that
atmospheric muon neutrinos undergo {\it large amplitude}
oscillations into another flavour. {\it Large amplitude oscillations imply
a large mixing angle between $\nu_{\mu}$ and another flavour. This is
exactly what is
provided for in the EPM through maximal $\nu_{\mu} \leftrightarrow \nu'_{\mu}$
mixing.} 
Furthermore, the anomalous zenith
angle dependence for multi-GeV $\mu$-like events reported by
SuperKamiokande \cite{kearns} provides strong independent evidence in favour 
of large
amplitude oscillations of $\nu_{\mu}$. The totality of atmospheric neutrino
data is well explained by $\nu_{\mu} \leftrightarrow \nu'_{\mu}$ 
oscillations with 
$\Delta m^2_{2+2-}$ in the approximate range
\begin{equation}
10^{-3} \stackrel{<}{\sim} \Delta m^2_{2+2-}/eV^2 \stackrel{<}{\sim}
10^{-2},
\label{mumuprime}
\end{equation}
where $\Delta m^2_{2+2-}$ is the
squared mass difference between $\nu_{2+}$ and $\nu_{2-}$ \cite{fvy1}. 
Note that if we restrict the discussion to two-flavour oscillations, then the 
present
data
allows only two choices: the atmospheric neutrino problem is solved either by
$\nu_{\mu} \leftrightarrow \nu_s$ oscillations (for which the EPM provides
an explicit theory) or
by $\nu_{\mu} \leftrightarrow \nu_{\tau}$ oscillations (see
Refs.\cite{fvy1,fvy2} for a
phenomenological study). In the future, these two alternatives may be 
experimentally
distinguished through neutral current effects \cite{vissani}, 
upward through-going
and stopping
muon data \cite{liu,liu2}, and long-baseline experiments.

The solar neutrino problem also provides strong evidence in favour of the
EPM neutrino sector. GALLEX \cite{gallex} and SAGE \cite{sage} 
observe a solar $\nu_e$ flux that is
close to half of that expected from the standard solar model when neutrino 
oscillations are absent. A $50\%$
$\nu_e$ flux reduction is exactly what is expected from the EPM due to
maximal $\nu_e \leftrightarrow \nu'_e$ oscillations for the mass range
\begin{equation}
10^{-10} \stackrel{<}{\sim} \Delta m^2_{1+1-}/eV^2 \stackrel{<}{\sim} 9
\times 10^{-4},
\label{eeprime}
\end{equation}
where the upper limit is required for consistency with the CHOOZ bound 
\cite{chooz}.
The other oscillation parameters are placed within the large region of
parameter space where intergenerational
solar $\nu_e$ oscillations are unimportant.
GALLEX and SAGE arguably provide the most unequivocal information regarding
the nature of the solar neutrino problem. There are two reasons for this:
First, theoretical
calculations of the expected event rates are the most robust. Second, 
both detectors have been calibrated with respect to a neutrino source of
known intensity. Kamiokande \cite{kamioka}, SuperKamiokande \cite{superk} 
and Homestake \cite{homestake} also provide
important information about solar neutrinos. All three of these experiments
report a significant deficit of solar neutrinos, leading to a qualitatively
consistent picture of solar $\nu_e$ depletion across the five experiments
(see Table \ref{expresults}).
Furthermore, because of the different energy thresholds of the experiments,
a comparison of their results provides information about the
energy-dependence of the solar neutrino flux depletion. Unfortunately, the
precise significance of the information obtained from Kamiokande,
SuperKamiokande and Homestake is less clear than for GALLEX and SAGE for
the following reasons: (i) Predictions for the Boron neutrino flux vary
significantly between different versions of the standard solar model,
mainly because of an uncertain $p + {^7}\text{Be} \to \gamma + {^8}\text{B}$ 
cross-section. The
precise value
of the Boron neutrino flux deficit is therefore not as well established as 
one would wish. (ii) The pioneering Homestake experiment is still the only
experiment that is especially sensitive to the mid-energy Beryllium
neutrinos. Other experiments are needed in order to confirm their result.
Fortunately, BOREXINO and the 
Iodine experiment will probe a similar part
of the spectrum in the near future. They will either confirm or disconfirm
the somewhat greater flux reduction reported by Homestake. 

The purpose of this paper is to study the range of possibilities for solar
neutrino flux depletion provided for by the EPM, and to determine the
implications of these possibilities for, in particular, the Sudbury
Neutrino Observatory (SNO) \cite{sno}. SNO will play a very important role 
in testing
the EPM because of its ability to distinguish between solar 
$\nu_e \leftrightarrow \nu_{\mu, \, \tau}$ oscillations and 
$\nu_e \leftrightarrow \nu'_{\alpha}$ oscillations
through its sensitivity to both charged and neutral current reactions.

It is important to understand that
the EPM supplies different solar neutrino outcomes in different regions of
parameter space. Hitherto \cite{fv95}, work within the EPM has focussed on the
simplest
and most characteristic possibility: First, parameters are chosen so that
intergenerational solar $\nu_e$ oscillations are unimportant. Maximal
$\nu_e \leftrightarrow \nu'_e$
oscillations in the range of Eq.(\ref{eeprime}) then lead to an {\it
energy-independent $50\%$ flux reduction} compared to no-oscillation
expectations. Furthermore, since $\nu'_e$ states are blind to the neutral
current, this case leads to an expectation that SNO will measure the
standard rate for charged current relative to neutral current events. This
case is in many ways the most attractive possibility within the EPM,
because it is extremely simple and because it most fully utilises the
predictive power of the EPM: a $50\%$ flux depletion is the direct result
of maximal $\nu_e \leftrightarrow \nu'_e$ mixing which in turn is the 
direct result of exact parity invariance. 

However, this case does not reproduce the greater
depletion of mid-energy neutrinos that is inferred from a comparison
between the Homestake rate and the other measured rates. In this paper, we
will explore regions of parameter space for the EPM that are different from
that considered above and hitherto. There are two principal motivations for
doing so. First, we want to identify those regions of parameter space that
can provide a better fit to the totality of solar neutrino data than can
the $50\%$ flux reduction region. Second, since this study will necessarily
involve solar $\nu_e$ oscillations into $\nu_{\mu}$, $\nu'_{\mu}$,
$\nu_{\tau}$ and $\nu'_{\tau}$ (as well as into $\nu'_e$), we will provide
interesting predictions for
the rate of charged current to neutral current events expected at SNO. Our
study will essentially be an exploration of the
Mikheyev--Smirnov--Wolfenstein (MSW) effect \cite{wolf,ms,review} within 
the EPM.  Note also
that maximal $\nu_e \leftrightarrow \nu'_e$ oscillations can provide an
energy-dependent flux reduction factor that fits all existing 
experiments well in the ``just-so'' regime \cite{justso},
\begin{equation}  
\label{justsomass}
\Delta m^2_{1+1-} \sim 5 \to 8 \times 10^{-11}eV^2.
\end{equation} 
Thus an experimental 
confirmation of an active--sterile just-so scenario
would also be consistent with the EPM for a tiny region of parameter space.

Before commencing the analysis we should comment that Big Bang Nucleosynthesis
poses a challenge for any model of light sterile neutrinos, due to the possible
excitation of excess degrees of freedom during the relevant cosmological 
epoch. 
Fortunately, and indeed remarkably, sterile neutrino models can generally
meet this challenge in full through the phenomenon of lepton asymmetry
generation by active--sterile or active--mirror oscillations. 
For a complete discussion, see Ref.\cite{bbn}. 

The rest of this paper is structured as follows: In Sec.\ \ref{sec2} 
we provide a
semi-quantitative analysis of the implications of the SuperKamiokande
measurement for the ratio of charged 
to neutral current event rates at SNO. Section \ref{sec3} 
deals with the 
mathematical formulation of the MSW effect on the EPM's underlying maximal
mixing framework.  
We present various MSW solutions 
to the solar neutrino problem within in EPM in Sec.\ \ref{sec4}.  
In Sec.\ \ref{sec5}, we discuss the  
implications of the solutions for the Boron neutrino energy spectrum and 
the Beryllium flux, and reexamine the SNO charged
to neutral current event rate, now constrained by five experiments.
We demonstrate that consistency with the LSND result can be attained 
in Sec.\ \ref{lsnd}.
We conclude in Sec.\ \ref{sec6}.

\section{From SuperKamiokande to SNO}
\label{sec2}

The Sudbury Neutrino Observatory (SNO) will determine whether solar
$\nu_e$'s oscillate into the other active flavours $\nu_{\mu}$ and
$\nu_{\tau}$ or into sterile flavours or, to some level of sensitivity,
into a combination of active and sterile flavours. {\it The last possibility is
the generic prediction of the EPM, because mirror neutrinos are sterile
with respect to ordinary weak interactions.} 

In this section, we will outline the various solar neutrino outcomes 
possible in the
EPM in different regions of parameter space. Our aim in this section 
is to estimate
the ratio of charged to neutral current event rates that SNO will measure 
if the EPM
is the correct theory of neutrino mixing. In this ``warm-up'' section, 
we will use as
little theoretical input as possible in order to not obscure, by the technical
complications of the MSW effect, the important phenomenological role 
played by the
characteristic maximal $\nu_{\alpha} \leftrightarrow \nu'_{\alpha}$ 
oscillations of
the EPM. Our minimal input will be: (i) that some type of MSW effect 
occurs (except
for Case A below); (ii) that averaged maximal vacuum 
$\nu_{\alpha} \leftrightarrow
\nu'_{\alpha}$ oscillations occur when the oscillation length is much 
less than an
astronomical unit; and, (iii)  that SuperKamiokande has measured the correct
depletion factor for Boron neutrinos.  We will revisit this issue in Sec.\
\ref{sec5}, armed with more detailed information about the required 
$\nu_e$ survival
probability.  This will allow us to further constrain our estimation 
of the ratio of
charged to neutrino current event rates expected at SNO.

\subsection{Case A: vacuum $\nu_e \leftrightarrow \nu'_e$ oscillations only}

Case A results from the parameter space region discussed in the
Introduction and in previous papers \cite{fv95}. Parameters are chosen 
so that the
only oscillation mode important for solar neutrinos is 
$\nu_e \leftrightarrow \nu'_e$ with
the oscillation length set by Eq.(\ref{eeprime}). If this case is correct,
SNO will measure the standard value for the ratio of charged current to
neutral current events. They should also, of course, confirm the
substantial depletion of Boron neutrinos reported by Kamiokande and
SuperKamiokande.

As a subcase of Case A, another possibility is maximal 
$\nu_e \leftrightarrow \nu'_e$ oscillations 
in the ``just-so'' regime.  The relevant parameters are given in 
Eq.(\ref{justsomass}).  This energy-dependent subcase is essentially an
active--sterile ``just-so'' scenario where the observed maximal mixing arises
from exact parity invariance.  In this picture, SNO will measure roughly the 
standard value for the ratio of charged to neutral current events.

\subsection{Cases employing the MSW effect}

Suppose intergenerational solar $\nu_e$ oscillations are now switched on by
choosing a different point in parameter space. In order to obtain
substantial intergenerational oscillations while simultaneously keeping the
relevant mixing angles small, the MSW mechanism must be invoked. We will
discuss the details of MSW transitions within the EPM in Sec. \ref{sec3}. 
For the
purposes of this section, we will merely suppose that MSW transitions exist
and deplete solar neutrinos in an appropriate energy-dependent fashion. 

The interesting issue for SNO is the flavour content of the solar neutrino
flux at the Earth. In general, MSW transitions will process some of the
solar $\nu_e$ flux into second and third generation neutrinos and mirror
neutrinos in the interior of the sun such that
\begin{equation}
\phi_0^{\odot}(e, E) = \phi^{\odot}(e, E) + \phi^{\odot}(e', E)
+ \phi^{\odot}(\mu, E) +
\phi^{\odot}(\mu', E) +
\phi^{\odot}(\tau, E) + \phi^{\odot}(\tau', E),
\label{sunflux}
\end{equation}
where $\phi_0^{\odot}(e, E)$ is the no-oscillation standard solar model
flux of
$\nu_e$ of energy $E$ at the surface of the sun, while
$\phi^{\odot}(\alpha, E)$ and
$\phi^{\odot}(\alpha', E)$ are the fluxes of
$\nu_{\alpha}$ and $\nu'_{\alpha}$, respectively, at the surface of
the sun. The
equality in Eq.(\ref{sunflux}) follows from flux conservation. The various
fluxes on the right hand side of Eq.(\ref{sunflux}) are given by
\begin{equation}
\phi^{\odot}(a, E) = P^{\odot}_{ea}(E) \phi^{\odot}_0(e, E),
\end{equation}
where $P^{\odot}_{ea}(E)$ is the matter-affected oscillation probability at
the surface
of the sun for $\nu_e \leftrightarrow \nu_a$ ($a = \alpha, \alpha'$). 
Probability
conservation requires that
\begin{equation}
1 = P^{\odot}_{ee}(E) + P^{\odot}_{ee'}(E)
+ P^{\odot}_{e\mu}(E) + P^{\odot}_{e\mu'}(E) 
+ P^{\odot}_{e\tau}(E) + P^{\odot}_{e\tau'}(E)
\label{probcon}
\end{equation}
for each value of $E$.

The crucial point can now be made: {\it between the sun and the Earth,
additional large amplitude vacuum oscillations will in general occur
between the maximally mixed 
standard plus mirror pairs.} Vacuum intergenerational oscillations will be
small given our assumption of small intergenerational mixing. Provided that
the oscillation lengths for $\nu_{\alpha} \leftrightarrow \nu'_{\alpha}$
are smaller than an astronomical unit (i.e., where the corresponding 
squared mass difference is 
$\stackrel{>}{\sim} 10^{-10}\ eV^2$), maximal vacuum oscillations will
induce
\begin{equation}
\phi^{\oplus}(\alpha, E) = \phi^{\oplus}(\alpha', E) =
\frac{\phi^{\odot}(\alpha, E) + \phi^{\odot}(\alpha', E)}{2\zeta},
\label{eqflux}
\end{equation}
where $\phi^{\oplus}$ denotes the flux {\it at the Earth}, and $\zeta$ is
a geometric factor due to the inverse square law. This is
expressed in terms of oscillation probabilities as
\begin{equation}
P^{\oplus}_{e\alpha}(E) = P^{\oplus}_{e\alpha'}(E) =
\frac{P^{\odot}_{e\alpha}(E) + P^{\odot}_{e\alpha'}(E)}{2},
\label{eqprobs}
\end{equation}
where the superscript $\oplus$ denotes oscillation probabilities at the
Earth. The different
possibilities for SNO now correspond to different oscillation length
regimes for $\nu_{\alpha} \leftrightarrow \nu'_{\alpha}$.

Because strong evidence now exists for an atmospheric neutrino anomaly, we
choose $\Delta m^2_{2+2-}$ in the range of Eq.(\ref{mumuprime}). This
means that Eqs.(\ref{eqflux}) and (\ref{eqprobs}) certainly hold for
$\alpha = \mu$. The
different cases discussed below correspond to the four generic
possibilities for $\Delta m^2_{1+1-}$ and $\Delta m^2_{3+3-}$.

\subsubsection{Case B}

Case B corresponds to the parameter choice
\begin{equation}
\Delta m^2_{1+1-},\ \Delta m^2_{3+3-} \stackrel{>}{\sim} 10^{-10}\ eV^2,
\label{Bol}
\end{equation}
so that Eqs.(\ref{eqflux}) and (\ref{eqprobs}) hold for 
$\alpha = e, \tau$ as well as for
$\alpha = \mu$. Combining Eqs.(\ref{sunflux}), (\ref{eqflux}) and
(\ref{Bol}) we see that the total flux of active neutrinos at the Earth
will be
\begin{equation}
\phi^{\oplus}(active, E) = \frac{\phi_0^{\oplus}(e, E)}{2} =
\frac{\phi_0^{\odot}(e, E)}{2\zeta},
\label{Bactiveflux}
\end{equation}
that is, exactly half of the no-oscillation $\nu_e$ flux. This $50\%$ flux
reduction is a direct result of the maximal mixing constraint following
from exact parity symmetry. This prediction implies that the SNO neutral
current rate will be $50\%$ of the no-oscillation expectation.

To quantify expectations for SNO, we consider the rates for charged current
and neutral current events given, respectively, by
\begin{eqnarray}
\Gamma_{CC} & = & \int^{\infty}_{E_0} P^{\oplus}_{ee}(E) 
\phi^{\oplus}_0(e, E)
\sigma_{CC}(E) dE,\nonumber\\
\Gamma_{NC} & = & \int^{\infty}_{E_0} [P^{\oplus}_{ee}(E) 
+ P^{\oplus}_{e\mu}(E) + P^{\oplus}_{e\tau}(E)] 
\phi_0^{\oplus}(e, E) \sigma_{NC}(E) dE,
\label{rates}
\end{eqnarray}
where $E_0$ is the energy threshold for SNO and $\sigma_{CC}(E)$
$[\sigma_{NC}(E)]$ is the charged [neutral] current cross-section.

According to Eqs.(\ref{probcon}), (\ref{eqprobs}), (\ref{Bol}) and
(\ref{rates}), we see that the charged to neutral current rate divided by
the no-oscillation expectation is given by
\begin{equation}
r_d \equiv
\frac{(\Gamma_{CC}/\Gamma_{NC})|_{osc}}{(\Gamma_{CC}/\Gamma_{NC})|_{0}}
= 2 \frac{\Gamma_{CC}|_{osc}}{\Gamma_{CC}|_0},
\label{rdB}
\end{equation}
where the characteristic factor of two is just another way of expressing
the $50\%$ flux reduction of the sum of active flavours.
SNO will measure
$\Gamma_{CC}|_{osc}$, while $\Gamma_{CC}|_0$ depends on Boron neutrino flux
predictions from the standard solar model. 

Equation (\ref{rdB}) is an exact result. In order to obtain a precise 
prediction for
$r_d$, the energy-dependent survival probability must be known. However, a good
estimation for $r_d$ can be obtained from the measured Boron neutrino flux at
SuperKamiokande, because the energy threshold of SNO is similar to that of
SuperKamiokande. The charged current event rate at SNO relative to the 
no-oscillation
expectation should be approximately equal to the analogous quantity
measured by SuperKamiokande. In order to use the SuperKamiokande measurement of
$\Omega_{SK}$, where
\begin{equation}
\Omega_{SK} \equiv \frac{\text{observed event rate}}{\text{no-oscillation event
rate}},
\end{equation}
we have to correct for the small contribution that neutral current induced 
$\nu_{\mu, \, \tau} e$ scattering
makes to it. Using Eqs.(\ref{Bactiveflux}) and (\ref{rdB}) together with 
the relation
$\sigma^{SK}(\nu_{\mu, \, \tau} e) \simeq
\frac{1}{6} \sigma^{SK}(\nu_e e)$ between the relevant cross-sections at
SuperKamiokande, we obtain
\begin{equation}
r_d \sim \frac{12\Omega_{SK} - 1}{5}
\end{equation}
for Case B.
Using the information in Table \ref{expresults}, and taking a $2\sigma$ limit
that incorporates both experimental and theoretical uncertainties, 
we get that
\begin{equation}
\Omega_{SK} \sim 0.25 \to 0.5.
\label{SKrange}
\end{equation}
The large range displayed here is mainly due to the significant theoretical
uncertainty in the Boron neutrino flux. [Note also that we have focussed on 
SSM-BP
(1995) only \cite{bp95}. Other SSM calculations yield significantly 
different Boron
neutrino fluxes \cite{ds}.] 
Using Eq.(\ref{SKrange}), we get
\begin{equation}
r_d \sim 0.4 \to 1.
\end{equation}
Note that for the upper extreme, where the Boron neutrino depletion is 
entirely due 
to averaged maximal $\nu_e \leftrightarrow \nu'_e$ oscillations, 
Case B becomes identical to Case A
for Boron neutrinos. 

By way of comparison, the standard 
$\nu_e \leftrightarrow \nu_{\mu, \, \tau}$
expectation is
\begin{equation}
r_d \sim \frac{6\Omega_{SK} - 1}{5} \sim 0.1 \to 0.4.
\end{equation}
(Note that SuperKamiokande data only have been used to obtain this estimate. 
Tighter
predictions are obtained when a survival probability consistent with all 
five solar
neutrino experiments is used.)
So, we see that the Case B range for $r_d$ covers all of the values between 
the
ranges for the standard $\nu_e \leftrightarrow \nu_{\mu, \, \tau}$ and
$\nu_e \leftrightarrow \nu_s$
solutions. A clear distinction between these three possibilities therefore 
seems to
be provided for by $r_d$, unless by bad luck the measured value turns out to 
be close
to either $0.4$ or $1$.

\subsubsection{Case C}

Case C is defined by the parameter choice
\begin{equation}
\Delta m^2_{1+1-} \stackrel{<}{\sim} 10^{-11}\ eV^2,\qquad 
\Delta m^2_{3+3-} \stackrel{>}{\sim} 10^{-10}\ eV^2,
\end{equation}
so that vacuum $\nu_e \leftrightarrow \nu'_e$ oscillations do not occur. 
In addition,
both direct transitions of $\nu_e$ to $\nu'_e$, and indirect transitions
via the second and third generation flavours, are negligible within the
sun. We can therefore set
\begin{equation}
P^{\odot}_{ee'}(E) = 0
\end{equation}
to a very good level of approximation. In this case, 
\begin{equation}
\phi^{\oplus}(active, E) = \frac{\phi_0^{\oplus}(e, E) 
+ \phi^{\oplus}(e, E)}{2}.
\label{Cactiveflux}
\end{equation}
The total flux of active flavours is larger than for Case B given the absence
of vacuum $\nu_e \leftrightarrow \nu'_e$ oscillations.

The ratio of charged to neutral current rates for this case relative 
to no-oscillation rates is given by
\begin{equation}
r_d = 2 \frac{\int^{\infty}_{E_0} P^{\oplus}_{ee}(E) \phi_0^{\oplus}(e, E)
\sigma_{CC}(E) dE}{\int^{\infty}_{E_0} [1 + P^{\oplus}_{ee}(E)]
\phi_0^{\oplus}(e, E) \sigma_{NC}(E) dE}
\frac{\int^{\infty}_{E_0} \phi_0^{\oplus}(e, E) \sigma_{NC}(E) dE}
{\int^{\infty}_{E_0} \phi_0^{\oplus}(e, E) \sigma_{CC}(E) dE},
\end{equation}
where an explicit expression for the energy-dependent $\nu_e$ survival
probability is required for an exact prediction. In principle, this has
to be done on a case by case basis. 

An approximate 
indication of the likely outcomes is obtained by neglecting the 
energy-dependence to
obtain
\begin{equation}
r_d \sim \frac{2\langle P^{\oplus}_{ee} \rangle}{1 + \langle P^{\oplus}_{ee}
\rangle},
\end{equation}
where $\langle \ldots \rangle$ denotes an average.
Taking the SuperKamiokande measurement of $\Omega_{SK}$, and correcting for
neutral current effects using Eq.(\ref{Cactiveflux}), we obtain
\begin{equation}
\langle P^{\oplus}_{ee} \rangle \sim \frac{12\Omega_{SK} - 1}{11}.
\end{equation}
For $\Omega_{SK}$ in the range of Eq.(\ref{SKrange}), this implies that
\begin{equation}
r_d \sim 0.3 \to 0.6.
\end{equation}
This case should be clearly distinguishable from Case A. It is not distinguishable
from Case B or from the standard MSW $\nu_e \to
\nu_{\mu, \,\tau}$ scenario on the basis of $r_d$ alone. We discuss this issue
further in Secs.\ \ref{sec5} and \ref{sec6}.

\subsubsection{Case D}

Case D corresponds to
\begin{equation}
\Delta m^2_{1+1-} \stackrel{>}{\sim} 10^{-10}\ eV^2,\qquad
\Delta m^2_{3+3-} \stackrel{<}{\sim} 10^{-11}\ eV^2.
\end{equation}
No especially interesting predictions can be made in this case without
further information. For
instance, if the MSW partners of $\nu_e$ are $\nu_{\mu}$ and
$\nu'_{\mu}$, then this case reduces to Case B. If, on the other hand, the
MSW partners of $\nu_e$ are
$\nu_{\tau}$ and $\nu'_{\tau}$, then, in the energy-independent
approximation,
\begin{equation}
r_d \sim \frac{\langle P^{\oplus}_{ee} \rangle}{\langle P^{\oplus}_{ee} \rangle +
\langle P^{\oplus}_{e\tau}\rangle}.
\end{equation}
For $\langle P^{\oplus}_{ee} \rangle = 0.25$, probability conservation at 
the surface
of the sun requires $0 < \langle P^{\oplus}_{e\tau} \rangle< 0.5$, leading 
to $r_d >
0.33$. The lower bound on $r_d$ increases with 
$\langle P^{\oplus}_{ee} \rangle$,
going to $1$ as $\langle P^{\oplus}_{ee} \rangle$ approaches $0.5$.

\subsubsection{Case E}

Finally, Case E corresponds to
\begin{equation}
\Delta m^2_{1+1-},\ \Delta m^2_{3+3-} \stackrel{<}{\sim} 10^{-11} eV^2.
\end{equation}
Again, more information is needed in this case in order to obtain predictions. 
If the
MSW partners of $\nu_e$ are $\nu_{\mu}$ and $\nu'_{\mu}$, then this case 
reduces
to Case C (and is also similar to the scheme analysed in Ref.\cite{liu}). 
If, on the
other hand, the MSW partners of $\nu_e$ are $\nu_{\tau}$ 
and $\nu'_{\tau}$, then this case is intermediate between the 
standard $\nu_e \leftrightarrow
\nu_{\mu, \, \tau}$  and $\nu_e \to \nu_s$ scenarios because vacuum 
oscillations play a
negligible role.

\section{Mathematical Formulation of the MSW effect in the EPM}
\label{sec3}

The EPM predicts an energy-independent $50\%$ reduction of the solar neutrino 
flux by maximal vacuum $\nu_e \leftrightarrow \nu'_e$ oscillations that is in
good quantitative agreement with experiments primarily sensitive to low-energy
and to high-energy neutrinos.  
The significantly lower event rate measured by 
Homestake, however, 
calls for further suppression of the mid-energy flux.
Preferential energy-dependent depletion can be achieved via the MSW mechanism 
by restricting the relevant intergenerational squared mass difference to 
\cite{ms}
\begin{equation}
\label{mswmass}
10^{-8} \stackrel{<}{\sim} \Delta m^2 /eV^2  \stackrel{<}{\sim} 10^{-4},
\end{equation}
and \cite{ms}, 
\begin{equation}
\label{mswmixing}
\sin^2 2 \eta  \stackrel{>}{\sim} 10^{-4},
\end{equation}
where $\eta$ parameterises the mixing between the corresponding neutrino 
states that take part in resonant conversion.

In the following analysis, a standard neutrino mass hierarchy, i.e.,
\begin{equation}
\label{eqhierarchy} 
m^2_{3\pm} > m^2_{2\pm} > m^2_{1\pm},
\end{equation}
is assumed such that the MSW
partners of $\nu_e$ are $\nu_{\mu}$ and $\nu'_{\mu}$.  Under the assumption 
of small intergenerational mixing, the contributions of $\nu_{\tau}$ 
and $\nu'_{\tau}$  towards the $\nu_e$ survival probability at the
Earth through vacuum
oscillations is negligibly small.  It then suffices to consider only the 
interactions between the first two generations, though the forthcoming 
mathematical treatments can be easily generalised to include the third
generation.  
The remaining two-generation system thus consists of four neutrino states,
where the transformation between the weak and mass/parity eigenstates 
is given by Eqs.(\ref{epmmixing1}) and (\ref{epmmixing2}).  Explicitly,
\begin{eqnarray}
\label{4numatrix}
\left( \begin{array}{c}
		\nu_e \\
		\nu'_e \\
		\nu_{\mu} \\
		\nu'_{\mu} \\
	\end{array} \right) 
= \frac{1}{\sqrt{2}} \left( \begin{array}{cccc}
			C_{\phi} & C_{\theta} & S_{\phi} & S_{\theta} \\
			-C_{\phi} & C_{\theta} & -S_{\phi} & S_{\theta} \\
			-S_{\phi} & -S_{\theta} & C_{\phi} & C_{\theta} \\
			S_{\phi} & -S_{\theta} & -C_{\phi} & C_{\theta} \\
			\end{array} \right)
			\left(\begin{array}{c}
					\nu_{1-} \\
					\nu_{1+} \\
					\nu_{2-} \\
					\nu_{2+} \\
			\end{array} \right),
\end{eqnarray}
where $\theta$ and $\phi$ parameterise the two $2 \times 2$ unitary matrices 
$U^{\pm}_{i\alpha}$ in Eq.(\ref{epmmixing2}) that are responsible for the 
respective mixing of
positive and negative parity eigenstates, and 
$-\frac{\pi}{2} \leq \theta, \, \phi \leq \frac{\pi}{2}$.  
Exact parity symmetry thereby 
reduces a nominally six-angle problem (if $CP$ is conserved) to a 
two-angle task.  (A generic $4 \times 4$ orthogonal matrix consists of six 
independent parameters.)

Note that, at this stage, we do not make any assumptions regarding the signs 
of $\Delta m^2_{1+1-}$ and $\Delta m^2_{2+2-}$, and Eq.(\ref{4numatrix}) does 
not imply in any way that $\nu_{1+}$ ($\nu_{2+}$) is heavier than $\nu_{1-}$
($\nu_{2-}$).  This is because, firstly, we have no prior reasons for doing 
so.  Secondly, since neutrino states of unlike parity do not mix in a gauge 
theoretic sense (i.e., the Lagrangian of the EPM in vacuum does not contain
parity-violating terms such as $m \bar{\nu}_{1+} \nu_{2-}$ for 
$\nu_{1+} \leftrightarrow \nu_{2-}$ \cite{fv95}), 
we would expect the {\it effective}
mixing of like-parity (such as $\nu_{1+} \leftrightarrow \nu_{2+}$)
and of unlike-parity eigenstates to experience different forms of matter
enhancement.  Ours being a two-angle problem renders the quantification of
this difference a relatively simple task.  Henceforth, we shall denote the
heavier (lighter) of $\nu_{1 \pm}$ and of $\nu_{2 \pm}$ as $\nu_{1h}$ and 
$\nu_{2h}$ ($\nu_{1 \ell}$ and $\nu_{2 \ell}$) such that
\begin{equation}
\label{eqb}
\left( \begin{array}{c}
	\nu_e \\
	\nu'_e \\
	\nu_{\mu} \\
	\nu'_{\mu} \\
	\end{array} \right)
= U \left( \begin{array}{c}
		\nu_{1 \ell} \\
		\nu_{1 h} \\
		\nu_{2 \ell} \\
		\nu_{2 h} \\
		\end{array} \right)
= \left( \begin{array}{cccc}
	U_{e 1 \ell} & U_{e 1h} & U_{e 2 \ell} & U_{e 2h} \\
	U_{e' 1 \ell} & U_{e'1h} & U_{e'2 \ell} & U_{e'2h} \\
	U_{\mu 1 \ell} & U_{\mu 1h} & U_{\mu 2 \ell} & U_{\mu 2h} \\
	U_{\mu' 1 \ell} & U_{\mu' 1h} & U_{\mu' 2 \ell} & U_{\mu' 2h} \\
	\end{array} \right)
	\left( \begin{array}{c}
		\nu_{1 \ell} \\
		\nu_{1 h} \\
		\nu_{2 \ell} \\
		\nu_{2 h} \\
		\end{array} \right),  
\end{equation}
in order to keep the analysis as general as possible.

The problem now becomes one of solving the Schr\"{o}dinger equation 
\cite{bil87}:
\begin{eqnarray}
\label{eqc}
i \frac{d}{dx} \left(\begin{array}{c}
			\nu_e \\
			\nu'_e \\
			\nu_{\mu} \\
			\nu'_{\mu} \\
		\end{array} \right)
& = &  {\cal H} \left(\begin{array}{c}
			\nu_e \\
			\nu'_e \\
			\nu_{\mu} \\
			\nu'_{\mu} \\
		\end{array} \right)
= \frac{1}{2E} (U {\cal M} U^{-1} + {\cal H}_{int})
		\left(\begin{array}{c}
			\nu_e \\
			\nu'_e \\
			\nu_{\mu} \\
			\nu'_{\mu} \\
		\end{array} \right)  \nonumber \\
& = &  \frac{1}{2E} \left[U \left( \begin{array}{cccc}
					m^2_{1 \ell} & 0 & 0 & 0 \\
					0 & m^2_{1h} & 0 & 0 \\
					0 & 0 & m^2_{2 \ell} & 0 \\
					0 & 0 & 0 & m^2 _{2h} \\
					\end{array} \right) U^{-1}
		+ \left( \begin{array}{cccc}
			A_e & 0 & 0 & 0 \\
			0 & 0 & 0 & 0 \\
			0 & 0 & A_{\mu} & 0 \\
			0 & 0 & 0 & 0 \\
			\end{array} \right) \right]
			\left(\begin{array}{c}
			\nu_e \\
			\nu'_e \\
			\nu_{\mu} \\
			\nu'_{\mu} \\
		\end{array} \right),
\end{eqnarray}
where $U$ is the mixing matrix in Eq.(\ref{eqb}), $E$ is the neutrino energy, 
and $m^2_{1 \ell}, \, m^2_{1h}, \, m^2_{2 \ell}$ and $m^2_{2h}$ are the squared
masses of the mass/parity eigenstates 
$\nu_{1 \ell}, \, \nu_{1h}, \, \nu_{2 \ell}$ and $\nu_{2h}$ respectively.  
The interaction terms for $\nu_e$ and $\nu_{\mu}$ 
($\nu'_e$ and $\nu'_{\mu}$ are inert) are
\begin{eqnarray}
\label{eqd}
A_e  & = & A_{CC} + A_{NC}, \nonumber \\
A_{\mu} & = & A_{NC},
\end{eqnarray}
where $CC$ stands for charged current, $NC$ for neutral current and
\begin{eqnarray}
\label{eqe}
A_{CC} & = & 2 \sqrt{2} G_{F} E N_e (x), \nonumber \\
A_{NC} & = & - \sqrt{2} G_{F} E N_n (x),
\end{eqnarray}
where $G_{F}$ is the Fermi constant, $N_e(x)$ the electron number density 
at position $x$ in the neutrino's path, and $N_n (x)$ the neutron number
density.  If $U^m$ is a density-dependent unitary transformation
that puts the total Hamiltonian ${\cal H}$ in an instantaneous mass basis 
$\nu^m_i$ such that
\begin{equation}
\label{eqf}
\nu_{\alpha} = \sum_{i} U^m_{\alpha i} \nu^m_i, \; 
		\alpha =  e, \, e', \, \mu, \, \mu', \;
		i =  1 \ell, \, 1h, \, 2 \ell, \, 2h,
\end{equation} 
the  Schr\"{o}dinger equation in Eq.(\ref{eqc}) can be rewritten as:
\begin{eqnarray}
\label{eqg}
i \frac{d}{dx} \left( \begin{array}{c}
			\nu^m_{1 \ell} \\
			\nu^m_{1h} \\
			\nu^m_{2 \ell} \\
			\nu^m_{2h} \\
			\end{array} \right)
& = & {\cal H}^m \left( \begin{array}{c}
			\nu^m_{1 \ell} \\
			\nu^m_{1h} \\
			\nu^m_{2 \ell} \\
			\nu^m_{2h} \\
			\end{array} \right)
= U^{m-1} ({\cal H} - i \frac{d}{dx}) U^m 
		\left( \begin{array}{c}
			\nu^m_{1 \ell} \\
			\nu^m_{1h} \\
			\nu^m_{2 \ell} \\
			\nu^m_{2h} \\
			\end{array} \right) \nonumber \\
& = & \left[ \frac{1}{2E} \left( \begin{array}{cccc}
				m^2_{1 \ell} (x) & 0 & 0 & 0 \\
				0 & m^2_{1h} (x) & 0 & 0 \\
				0 & 0 & m^2_{2 \ell}(x) & 0 \\
				0 & 0 & 0 & m^2_{2h}(x) \\
				\end{array} \right)
- U^{m-1} i \frac{d}{dx} U^m \right] \left( \begin{array}{c}
					\nu^m_{1 \ell} \\
					\nu^m_{1h} \\
					\nu^m_{2 \ell} \\
					\nu^m_{2h} \\
					\end{array} \right),
\end{eqnarray}
where $m^2_{1 \ell} (x), \, m^2_{1h} (x), \, m^2_{2 \ell} (x)$ and 
$m^2_{2h} (x)$ are the squared mass eigenvalues of the instantaneous mass 
eigenstates
$\nu^m_{1 \ell}, \, \nu^m_{1h}, \, \nu^m_{2 \ell}$ and $\nu^m_{2h}$ 
respectively.  Given the initial conditions
\begin{equation}
\nu_{e} (x_i) = 1, \; \nu'_e (x_i) = \nu_{\mu} (x_i) = \nu'_{\mu} (x_i) = 0, 
\end{equation}
where $x_i$ is the $\nu_e$ production position,
the probability that a $\nu_e$ produced in the sun will be 
detected on Earth is expressed as 
\begin{equation}
\label{eqh}
P^{\oplus}_{ee} (E) =  \left| \sum_{i, j} U^m_{ei} (x_i) 
		\exp \left[ - i \int^{x_f}_{x_i} {\cal H}^m_{ij} dx \right]
			U_{ej} \right|^2,
\end{equation}
where $i, \, j = 1 \ell, \, 1h, \, 2 \ell, \, 2h$, the exponential of the 
integral from $x_i$ to the detection position $x_f$ is the solution
to Eq.(\ref{eqg}), $U^m (x_i)$ is the density-dependent mixing matrix at the 
point of $\nu_e$ production and we have chosen $U$ real (assuming that $CP$ is
conserved).  The term 
$\exp \left[-i \int^{x_f}_{x_i} {\cal H}^m_{ij} dx \right]$
represents
the amplitude of a transition from $\nu^m_i$  to $\nu^m_j$ and vice versa.

For future reference, note that for a $2\nu$ system, 
after phase-averaging, 
the $\nu_e$ survival probability is given by \cite{parke}
\begin{equation}
\label{eqi}
P^{\oplus}_{ee} (E)|_{2 \nu} = \frac{1}{2} + (\frac{1}{2} - P_R) 
		\cos 2 \eta_m (x_i)\cos 2 \eta
\equiv P_{2 \nu MSW},
\end{equation}
where $\eta$ is the vacuum mixing angle, $\eta_m (x_i)$ the matter mixing
angle at the $\nu_e$ production position, and $P_R$ the level-crossing 
probability evaluated at resonance.  [A common practice is to multiply the 
$P_R$ term by a step function, $\theta (E - E_A)$, where $E_A$ is the minimum
energy a neutrino produced at $x_i$ must possess for a 
resonance to take place
inside the sun \cite{kuo87}.  
It shall be omitted for the convenience of 
typesetting.]  We shall refer to this $2\nu$ survival
probability as the MSW transition probability $P_{2 \nu MSW}$.  
Bearing in mind that the same expression 
can be obtained by considering probabilities instead of amplitudes, we shall
adopt the same classical attitude for the rest of the analysis.  

An analysis involving two effectively sterile neutrinos approximately 
maximally mixed with $\nu_e$ and $\nu_{\mu}$ respectively was carried 
out in Ref. \cite{giunti}.  The ``intragenerational'' mass differences were 
assumed to be much smaller than the intergenerational
 mass difference, the latter of which was responsible for a 
$\nu_e \leftrightarrow \nu_{\mu}$ MSW resonance.  Our case differs in that 
the explanation of the atmospheric neutrino anomaly by maximal vacuum 
$\nu_{\mu} \leftrightarrow \nu'_{\mu}$ oscillations requires the squared mass 
difference between $\nu_{2+}$ and $\nu_{2-}$ to lie within the range of
Eq.(\ref{mumuprime}).  These masses are much larger than the MSW masses 
in Eq.(\ref{mswmass}),
i.e.,
\begin{equation}
\label{eqj}
\Delta m^2_{2+2-} \gg \Delta m^2_{21},
\end{equation}
where $\Delta m^2_{21} = m^2_{2 \ell} - m^2_{1h}$.  On the other hand, the 
squared mass difference between $\nu_{1 \ell}$ and $\nu_{1h}$  
is constrained only by an experimental upper bound of 
$9 \times 10^{-4} eV^2$ from Eq.(\ref{eeprime}).  
Thus, analytically {\it a priori} well-approximated 
MSW solutions 
exist for three distinct neutrino mass hierarchies, which we shall denote as 
Cases B1, B2, and C respectively: \linebreak
\vspace{1mm}
Case B1: $\Delta m^2_{1+1-}, \, \Delta m^2_{2+2-} \gg \Delta m^2_{21}$, \\
\vspace{1mm}
Case B2: $\Delta m^2_{2+2-} \gg \Delta m^2_{21} \gg \Delta m^2_{1+1-} 
		\stackrel{>}{\sim} 10^{-10} eV^2$, \\
\vspace{1mm}
Case C: $\Delta m^2_{2+2-} \gg \Delta m^2_{21} \gg \Delta m^2_{1+1-}, \;
		\Delta m^2_{1+1-} \stackrel{<}{\sim} 10^{-11} eV^2$. \\
These somewhat playful labels are chosen for consistency with Sec. \ref{sec2}, 
i.e.,
the generic predictions for SNO for Case B in Sec. \ref{sec2} 
hold for both neutrino mass 
hierarchies defined in Cases B1 and B2, and similarly for Case C. 
In the following subsections, we shall derive the $\nu_e$ survival 
probability for each case.

\subsection{Case B1}

Case B1 assumes the following neutrino mass hierarchy:
\begin{equation}
\label{eqb101}
\Delta m^2_{1+1-}, \Delta m^2_{2+2-} \gg \Delta m^2_{21}.
\end{equation}
In this analysis, we
make one further assumption that
\begin{equation}
\label{eqb102}
\Delta m^2_{1+1-} \sim 10^{-3} eV^2
\end{equation}
for simplicity.  The squared masses of the instantaneous mass eigenstates
 for the relevant solar
densities are shown in Fig. \ref{lsl}.

We identify the point $R$, at which $\nu^m_{1h}$ and $\nu^m_{2 \ell}$ 
almost 
cross, as an intergenerational MSW resonance.  Large 
$\Delta m^2_{1+1-}$ and $\Delta m^2_{2+2-}$
(compared with $\Delta m^2_{21}$) ensure that 
$\nu_e \leftrightarrow \nu'_e$  and 
$\nu_{\mu} \leftrightarrow \nu'_{\mu}$ oscillations remain close to maximal 
in its vicinity.  Consequently,
matter effects are 
most strongly felt by $\nu_{1h}$ and $\nu_{2 \ell}$, leading to the 
instantaneous mass eigenstates $\nu^m_{1h}$ and $\nu^m_{2 \ell}$ bearing
little resemblance to their vacuum counterparts.  The evolution of 
$\nu_{1 \ell}$
and $\nu_{2h}$ near $R$, on the other hand, is only slightly affected by
matter, so that 
\begin{eqnarray}
\label{eqb103}
\nu^m_{1 \ell} & \simeq & \nu_{1 \ell}, \nonumber \\
\nu^m_{2h} & \simeq &\nu_{2h}.
\end{eqnarray}
Provided Eq.(\ref{eqb102}) is satisfied, Eq.(\ref{eqb103}) will continue 
to hold at densities $\rho \gg \rho_R$ in the sun.  Thus, to a very good level
of approximation,
\begin{eqnarray}
\label{eqb104}
U^m_{e 1 \ell} (x_i) & = &  U_{e 1 \ell}, \nonumber \\
U^m_{e 2h} (x_i) & = & U_{e 2h},
\end{eqnarray}
which are virtually density-independent.  [Note that if $\Delta m^2_{1+1-}$
satisfies Eq.(\ref{eqb101}) but is, at the same time, sufficiently small (say,
$\sim 10^{-5} eV^2$), matter effects can cause $\nu_e$ and $\nu'_e$ to depart 
from their mutual maximal mixing at $\rho \gg \rho_R$.  $U^m_{e 1 \ell} (x_i)$
becomes density-dependent and may be drastically different from its vacuum 
counterpart.  The quantification of this effect is relatively simple.  
For our purposes, however, we shall not consider it here.]  Thus, 
$\nu^m_{1 \ell}$ ($\sim \nu_{1 \ell}$) and $\nu^m_{2h}$ ($\sim \nu_{2h}$) 
decouple from the system and evolve 
adiabatically such that after phase-averaging, we may write, 
following the procedures in Ref.\cite{bgkp}, 
the $\nu_e$ survival probability as
\begin{eqnarray}
\label{eqb105}
P^{\oplus}_{ee} (E) & = & |U_{e 1 \ell}|^4 + |U_{e 2h}|^4
			+ \left| \sum_{i, \, j = 1h, \, 2 \ell}
			U^m_{ei} (x_i)
		\exp \left[- i \int^{x_f}_{x_i} {\cal H}^m_{ij} dx \right] 
			U_{ej} \right|^2 \nonumber \\
& = & |U_{e 1 \ell}|^4 + |U_{e 2h}|^4
	+ (1 - |U_{e 1 \ell}|^2 - |U_{e 2h}|^2)^2 P_{2 \nu MSW},
\end{eqnarray}
where
\begin{equation}
\label{eqb106}
P_{2 \nu MSW} =  \left| \sum_{i, \, j = 1h, \, 2 \ell}
			\frac{U^m_{ei} (x_i)}
			{\sqrt{U^{m2}_{e1h} + U^{m2}_{e2 \ell}}}
		\exp \left[- i \int^{x_f}_{x_i} {\cal H}^m_{ij} dx \right] 
			\frac{U_{ej}}{\sqrt{U^2_{e1h} + U^2_{e2 \ell}}}
			 \right|^2,
\end{equation}
and $U^{m2}_{e 1h} + U^{m2}_{e 2 \ell} = U^2_{e1h} + U^2_{e 2 \ell}$ by
unitarity.
The physical interpretation of Eqs.(\ref{eqb105}) and (\ref{eqb106}) follows
simply.  At well above the resonance density, a fraction $|U^m_{e1 \ell} (x_i)|^2$
($\sim |U_{e1 \ell}|^2$) and another $|U^m_{e2h} (x_i)|^2$ 
($\sim |U_{e2h}|^2$) 
of the $\nu_e$'s produced at $x_i$ populate the instantaneous mass eigenstates 
$\nu^m_{1 \ell}$ ($\sim \nu_{1 \ell} \sim \frac{1}{2}$) and 
$\nu^m_{2h}$ ($\sim \nu_{2h} \sim 0$)  
respectively.  The remaining 
$(1 - |U_{e1 \ell}|^2 - |U_{e2h}|^2 \sim \frac{1}{2})$ is
distributed in $\nu^m_{1h}$ and $\nu^m_{2 \ell}$ in a density-dependent
ratio, $\cos^2 2 \eta_m : \sin^2 2 \eta_m$, where
\begin{equation}
\label{eqb107}
\cos \eta_m = \frac{U^m_{e1h}}{\sqrt{U^{m2}_{e1h} + U^{m2}_{e2 \ell}}}, \;
\sin \eta_m = \frac{U^m_{e2 \ell}}{\sqrt{U^{m2}_{e1h} + U^{m2}_{e2 \ell}}},
\end{equation}
and $\eta_m \sim \frac{\pi}{2}$ for $\rho \gg \rho_R$.
So the $\nu_e$'s produced in the region $\rho \gg \rho_R$ near the 
centre of the sun populate $\nu^m_{1 \ell}, \, \nu^m_{1h}, \, \nu^m_{2 \ell}$
and $\nu^m_{2h}$ in the approximate ratio 
$\frac{1}{2} : 0 \frac{1}{2} : 0$.  The half residing in $\nu^m_{2 \ell}$
participate in resonant conversion at $R$, while the other half in 
$\nu^m_{1 \ell}$ ($\sim \nu_{1 \ell}$) propagate adiabatically to the surface 
of the sun without passing through a resonance. 

To study the intergenerational MSW resonance, we may treat the 
$\nu_{1h} \leftrightarrow \nu_{2 \ell}$ subsystem as forming two orthogonal
pseudo-weak eigenstates, $\nu_a$ and $\nu_b$, that convert resonantly into 
each other at $R$, i.e.,
\begin{equation}
\label{eqb108}
\left( \begin{array}{c}
	\nu_{1 \ell} \\
	\nu_{a} \\
	\nu_{b} \\
	\nu_{2h} \\
	\end{array} \right)
= {\cal R}(\eta) \left( \begin{array}{c}
		\nu_{1 \ell} \\
		\nu_{1h} \\
		\nu_{2 \ell} \\
		\nu_{2h} \\
		\end{array} \right)
= \left( \begin{array}{cccc}
	1 & 0 & 0 & 0 \\
	0 & \cos \eta & \sin \eta & 0 \\
	0 & - \sin \eta & \cos \eta & 0 \\
	0 & 0 & 0 & 1 \\
	\end{array} \right)
	\left( \begin{array}{c}
		\nu_{1 \ell} \\
		\nu_{1h} \\
		\nu_{2 \ell} \\
		\nu_{2h} \\
		\end{array} \right),
\end{equation}
where
\begin{equation}
\label{eqb109}
\cos \eta = \frac{U_{e1h}}{\sqrt{U^2_{e1h} + U^2_{e2 \ell}}}, \;
\sin \eta = \frac{U_{e2 \ell}}{\sqrt{U^2_{e1h} + U^2_{e2 \ell}}},
\end{equation}
are the vacuum counterparts of the parameters in Eq.(\ref{eqb107}).  With
this parameterisation, we may rewrite the mixing matrix $U$ as
\begin{equation}
\label{eqb110}
U = {\cal T}(others) {\cal R}(\eta),
\end{equation}
where ${\cal T}(others)$ is a unitary matrix responsible for other mixing 
modes.  Since $\nu_{1h} \leftrightarrow \nu_{2 \ell}$ is the only 
matter-enhanced mixing mode, with the matter mixing angle $\eta_m$ taking 
on the maximal value $\frac{\pi}{4}$ at resonance, we may approximate $U^m$ as
\begin{eqnarray}
\label{eqb111}
U^m & = & {\cal T}(others) {\cal R}(\eta_m) \nonumber \\
& = & U {\cal R}^{-1}(\eta) {\cal R}(\eta_m),
\end{eqnarray}
in the same manner that is adopted in the study of $3\nu$ 
schemes \cite{kuo87,kuo86}.  Equation ({\ref{eqb111}) then
allows us to recast the integrand ${\cal H}^m_{ij}$ in Eq.(\ref{eqb106})
into
\begin{eqnarray}
\label{eqb112}
{\cal H}^m & = & U^{m-1} ({\cal H} - i \frac{d}{dx}) U^m \nonumber \\ 
& = & {\cal R}^{-1} (\eta_m) ({\cal H}_{eff} - i \frac{d}{dx}){\cal R}(\eta_m),
\end{eqnarray}
where
\begin{eqnarray}
\label{eqb113}
{\cal H}_{eff} & = & {\cal R}(\eta) U^{-1} {\cal H} U {\cal R}^{-1} (\eta)
\nonumber \\
& = & \frac{1}{2E} [{\cal R}(\eta) {\cal M} {\cal R}^{-1} (\eta) +
{\cal R}(\eta) U^{-1} {\cal H}_{int} U {\cal R}^{-1} (\eta)],
\end{eqnarray}
by Eqs.(\ref{eqc}) and (\ref{eqg}).  
The $2 \times 2$ submatrix 
${\cal H}_{effij}$, where 
$i, \, j = 1h, \, 2 \ell$ (i.e., the $2-3$ sector of ${\cal H}_{eff}$), is 
thus the effective 
Hamiltonian that governs the
evolution of the pseudo-weak eigenstates $\nu_a$ and $\nu_b$.  Subtracting a
common phase, ${\cal H}_{effij}$ takes on the form
\begin{equation}
\label{eqb114}
{\cal H}_{effij} \doteq \frac{1}{4E}
		\left( \begin{array}{cc}
		- \Delta m^2_{21} \cos 2 \eta + A_{eff}
		& \Delta m^2_{21} \sin 2 \eta + A_{ind} \\
		\Delta m^2_{21} \sin 2 \eta + A_{ind}
		& \Delta m^2_{21} \cos 2 \eta - A_{eff} \\
		\end{array} \right),
\end{equation} 
where $A_{eff}$, the effective density, is
\begin{eqnarray}
\label{eqb115}
A_{eff} & = & A_{CC} (U^2_{e1h} + U^2_{e2 \ell}) \nonumber \\
	& + & A_{NC} [\cos 2 \eta (U^2_{e1h} - U^2_{e2 \ell} 
	+ U^2_{\mu 1h} - U^2_{\mu 2 \ell})
	+ 2 \sin 2 \eta (U_{e1h} U_{e2 \ell} + U_{\mu 1h} U_{\mu 2 \ell})],
\end{eqnarray}
and   
\begin{equation}
\label{eqb116}
A_{ind} = A_{NC} [2 \cos 2 \eta (U_{e1h} U_{e2 \ell} + 
			U_{\mu 1h} U_{\mu 2 \ell})
		- \sin 2 \eta (U^2_{e1h} - U^2_{e 2 \ell} 
		+ U^2_{\mu 1h} - U^2_{\mu 2 \ell})],
\end{equation}
where the subscript $ind$ stands for induced.  The physical significance of
this term will be discussed in due course.  The quantities $A_{eff}$ 
and $A_{ind}$
evaluated for various combinations of $\nu_{1h}$ and $\nu_{2 \ell}$ in the
EPM are shown in Table \ref{bonetable}.  
The effective Hamiltonian ${\cal H}_{effij}$ is 
analogous to that for a standard $2\nu$ system.  The solution to
\begin{equation}
i \frac{d}{dx} \left( \begin{array}{c}
			\nu_a \\
			\nu_b \\
			\end{array} \right)
= {\cal H}_{effij} \left( \begin{array}{c}
			\nu_a \\
			\nu_b \\
			\end{array} \right)
\end{equation}
will thus give us the term $P_{2 \nu MSW}$ that is equivalent to 
the MSW transition probability given by Eq.(\ref{eqi}), 
with a squared mass difference of 
$\Delta m^2_{21}$ and mixing angle $\eta$ defined in Eq.(\ref{eqb109}) in a 
medium of effective density $A_{eff}$ given by Eq.(\ref{eqb115}) (plus some
non-standard features to be discussed).

We now look at the $2 \nu$ subsystem more closely.  
In general, the effective density $A_{eff}$ in Eq.(\ref{eqb115}) contains 
both charged and neutral current interaction
terms, though the latter's contribution is negligible if 
intergenerational mixing is small, i.e.,
\begin{equation}
\label{eqb117}
A_{eff} \approx \frac{1}{2} A_{CC},
\end{equation}
for $|U_{e1h}|,\, |U_{\mu 2 \ell}| \sim \frac{1}{\sqrt{2}}$ and 
$|U_{e2 \ell}|, \, |U_{\mu 1h}| \sim 0$.  Equation (\ref{eqb117}), in
turn, 
supports an approximate resonance condition given by
\begin{equation}
\label{eqb118}
A_{CC} \approx 2 \Delta m^2_{21} \cos 2 \eta,
\end{equation}
which may be rearranged into a more illuminating form:
\begin{equation}
\label{eqb119}
E_A \approx \frac{\Delta m^2_{21} \cos 2 \eta}{\sqrt{2} G_F N_e(x_i)},
\end{equation}
where the subscript $A$ stands for adiabatic and $N_e(x_i)$ is the electron
number density at the $\nu_e$ production position.  The quantity $E_A$ 
determines the 
location of the adiabatic edge of the MSW transition probability in the limit
of small intergenerational mixing, such that for
all neutrinos produced at $x_i$, only the ones with energy $E > E_A$ will be
resonantly converted into other species.  Comparing Eq.(\ref{eqb119}) to
its counterpart in the standard $\nu_e \leftrightarrow \nu_{\mu, \, \tau}$
scenario 
(where $A_{eff} = A_{CC}$), our naturally smaller effective density
automatically puts the adiabatic edge at twice the energy of the latter for a
given $\Delta m^2_{21}$ and $\eta$.
Equations (\ref{eqb117}), (\ref{eqb118}) and (\ref{eqb119}) are exact and 
{\it independent} of 
intergenerational mixing (provided it is nonzero), according to 
Table \ref{bonetable} if (i) $\nu_{1h}$ and 
$\nu_{2 \ell}$ are both positive or both negative parity eigenstates,
or (ii) $\theta = \phi$ for all possible 
combinations of $\nu_{1h}$ and $\nu_{2 \ell}$.  Nonetheless, while we are not 
considering large $\theta$ and $\phi$ cases, Eq.(\ref{eqb119}) will locate the
adiabatic edge with sufficient accuracy regardless of the exact identities of 
$\nu_{1h}$ and $\nu_{2 \ell}$ for the present 
analysis in the limit of small intergenerational mixing.  Taking 
$|U_{e1 \ell}| \sim \frac{1}{\sqrt{2}}$ and 
$|U_{e2h}| \sim 0$, the $\nu_e$ survival probability for Case B1 in 
Eq.(\ref{eqb105}) is well approximated by
\begin{equation}
\label{eqb119a}
P^{\oplus}_{ee} (E) \approx \frac{1}{4} + \frac{1}{4} P_{2 \nu MSW}.
\end{equation}   
Equation (\ref{eqb119a}) is plotted in Fig. \ref{lslprob}, juxtaposed 
with the respective
survival probabilities for the standard 
$\nu_e \leftrightarrow \nu_{\mu, \, \tau}$ and 
$\nu_e \leftrightarrow \nu_s$ cases evaluated for  the same oscillation
parameters 
for comparison.

The second non-standard feature is the presence of density-dependent 
terms, $A_{ind}$, in the 
off-diagonal elements of the effective Hamiltonian in Eq.(\ref{eqb114}), 
representing 
some form of matter-induced mixing similar to that discussed in 
Ref.\cite{liu}. 
This matter-induced mixing manifests itself primarily in the non-adiabatic 
high-energy end of the MSW transition probability.  
Explicitly, if we write the effective
Hamiltonian as 
\begin{equation}
\label{eqb120}
{\cal H}_{eff} = \left( \begin{array}{cc}
			- \xi (x) & \kappa (x) \\
			\kappa (x) & \xi (x) \\
			\end{array} \right), 
\end{equation}
assuming a linear density profile, the level-crossing probability
may be written as \cite{lz,petcov}
\begin{equation}
\label{eqb121}
P_R = \exp \left[ -\frac{\pi}{2} \gamma_R \right],
\end{equation}
where the adiabaticity parameter $\gamma_R$ is
\begin{equation}
\label{eqb122}
\gamma_R = \left. 2\frac{\kappa^2(x)}{\left| \frac{d \xi(x)}{dx} \right|} 
			 \right|_{x = x_R},
\end{equation}
evaluated at resonance;  $P_R$'s 
dependence on the matter-induced mixing term $A_{ind}$ is obvious.

In the context of the EPM, the extent to which matter-induced mixing affects 
the non-adiabatic conversion of solar neutrinos depends largely on the 
identities of $\nu_{1h}$ and $\nu_{2 \ell}$.  In particular, the mixing of
like-parity and of unlike-parity eigenstates receive considerably different
forms of matter enhancement.  With reference to Table \ref{bonetable}, if 
$\nu_{1h}$ and $\nu_{2 \ell}$  
are like-parity eigenstates, $A_{ind}$ vanishes
exactly, leaving behind in Eq.(\ref{eqb114}) the standard vacuum parameters 
$\Delta m^2_{21}$ and $\eta$, where 
$\eta$ is now replaced with $\theta$ or $\phi$ for 
$\nu_{1+} \leftrightarrow \nu_{2+}$ and $\nu_{1-} \leftrightarrow \nu_{2-}$ 
respectively.  Thus, 
parity-conserving, direct mixing modes are enhanced naturally by matter 
effects in a familiar resonant fashion.  In addition, the resonant enhancement
of one mixing mode is completely independent of the other, that is, if the 
resonant mode is $\nu_{1+} \leftrightarrow \nu_{2+}$ where $\theta$ is the 
mixing angle responsible, $\phi$ does not enter the scene.  

In the EPM, the {\it apparent} mixing of unlike-parity eigenstates in vacuum
is an observational effect due to mixing through other parity-conserving
channels.  In matter, {\it apparent} parity-violating mixing is, 
to some extent,
conjured up by matter, as suggested by the general presence of an 
$A_{ind}$ term for the matter-enhanced effective mixing of $\nu_{1-}
\leftrightarrow \nu_{2+}$ and $\nu_{1+} \leftrightarrow \nu_{2-}$
respectively.  
Furthermore, the strength
of this matter-induced mixing is dependent on the relative amplitude of 
the $\theta$ and $\phi$ modes.  An inspection of Table \ref{bonetable} 
reveals that, 
depending on the sign of 
$\sin (\theta - \phi)$, matter-induced mixing may enhance or suppress the 
non-adiabatic conversion of $\nu_e$ by decreasing or increasing respectively 
the matter 
oscillation length at resonance.  The magnitude of this matter-induced mixing 
is, in part, controlled by the neutron density at resonance but is most 
severe when $\theta$ and $\phi$ differ significantly.  Consider the case of
effective $\nu_{1-} \leftrightarrow \nu_{2+}$ mixing.  The contribution from
matter-induced mixing relative to vacuum mixing is represented by the ratio
\begin{equation}
\label{eqb123}
\Lambda = \left| \frac{A_{NC} \cos 2 \eta \sin (\theta - \phi)}
			{\Delta m^2_{21} \sin 2 \eta} \right|_{x = x_R},
\end{equation}
evaluated at resonance.  Using the resonance condition in 
Eq.(\ref{eqb118}), we find that
\begin{equation}
\label{eqb124}
\Lambda  \cong  \frac{N_n(x_R)}{N_e(x_R)} 
		\left|\frac{\cos^2 2 \eta \sin (\theta - \phi)}
			{\sin 2 \eta} \right|  
 \cong  \frac{1}{2} \frac{N_n(x_R)}{N_e(x_R)}
			\left| 1 - \frac{\tan \phi}{\tan \theta} \right|,
\end{equation}
by various relations in Table \ref{bonetable} to first order in 
$\sin \eta$.  
Matter-induced and vacuum 
mixing are comparable if $\Lambda \approx 1$.  Given that the electron 
number density is some two
to six times the neutron number density in the interior of 
the sun \cite{bp95}, this 
corresponds to 
\begin{equation}
\label{eqb125a}
\frac{\tan \phi}{\tan \theta}  <  2 \frac{N_e(x_R)}{N_n(x_R)} + 1  =  
5 \to 13,
\end{equation}
if $\theta$ and $\phi$ are in the same quadrant, or
\begin{equation}
\label{eqb125b}
\frac{\tan \phi}{\tan \theta}  <  1 - 2\frac{N_e(x_R)}{N_n(x_R)}  =  
-11 \to -3,
\end{equation}
if $\theta$ and $\phi$ are in different quadrants, 
in order for matter-induced mixing to be recessive.  The most extreme 
scenario is when the $\theta$ mode is completely absent, such that 
$\sin \eta = 0$ and matter-induced mixing completely dominates.  The 
consequential shift
of the adiabatic edge in the MSW transition probability is negligible.  On the 
non-adiabatic side, assuming a linear density profile, the level-crossing 
probability
is determined by
\begin{eqnarray}
\label{eqb126}
P_R & = & \left. \exp \left[ -\frac{\pi}{4} 
	\frac{\Delta m^2_{21}}{E} 
	\frac{\left(\frac{|A_{NC}|}{\Delta m^2_{21}} \sin \phi \right)^2}
		{\left| \frac{1}{A_{eff}} 
		\frac{d A_{eff}}{dx} \right|} \right] \right|_{x = x_R}
	\nonumber \\
& = & \left. \exp \left[ - \frac{\pi}{4} 
	\frac{\Delta m^2_{21}}{E} 
	\frac{\sin^2 2 \phi}
		{\left|\frac{1}{A_{eff}} 
		\frac{d A_{eff}}{dx} \right|}
	\left( \frac{|A_{NC}|}{2 \Delta m^2_{21} \cos \phi} \right)^2 \right]
	\right|_{x = x_R}.
\end{eqnarray} 
Comparing this with its counterpart in the case where 
$\theta = \phi \equiv \varphi$ (such that 
$A_{ind}$ vanishes exactly, see Table \ref{bonetable}), that is,
\begin{equation}
\label{eqb127}
P_R= \left. \exp \left[ - \frac{\pi}{4}
	\frac{\Delta m^2_{21}}{E}
	\frac{\sin^2 2 \varphi}{\cos 2 \varphi}
	\frac{1}{\left| \frac{1}{A_{eff}} 
	\frac{d A_{eff}}{dx} \right|} \right] \right|_{x = x_R},
\end{equation}
we observe that approximate agreement between Eqs.(\ref{eqb126}) and
(\ref{eqb127}) in the small intergenerational mixing limit requires (for the
same $\Delta m^2_{21}$),
\begin{eqnarray}
\label{eqb128}
\sin^2 2 \phi & \cong &	\left(\frac{2 \Delta m^2_{21}}{|A_{NC}|} \right)^2 
			\sin^2 2 \varphi
 	\cong \left( \frac{2 N_e(x_R)}{N_n(x_R)} \right)^2 
		\sin^2 2 \varphi \nonumber \\
	& \cong & (16 \to 144) \sin^2 2 \varphi.
\end{eqnarray}
Thus, if we fit the $\nu_e$ survival probability in Eq.(\ref{eqb119a}) to
experimental data for the $\theta = \phi$ case (so that the 
identities of $\nu_{1h}$ and $\nu_{2 \ell}$ do not matter), we know 
automatically from Eqs.(\ref{eqb117}) to (\ref{eqb119}) that approximately 
the same 
$\Delta m^2_{21}$ will account for $\nu_{1-} \leftrightarrow \nu_{2+}$
with $\theta = 0$, while the mixing required, $\sin^2 2 \phi$, is some 
$16$ to $144$ times that for $\theta = \phi$, under the assumption of small
intergenerational mixing according to Eq.(\ref{eqb128}).

On the other hand, if $\phi$ is set to zero and $\theta$ allowed to vary for 
$\nu_{1-} \leftrightarrow \nu_{2+}$, we see from Table \ref{bonetable} 
that both vacuum
and matter-induced mixing contribute to the level-crossing probability, i.e.,
\begin{eqnarray}
\label{eqb129}
P_R & = & \left. \exp \left[ - \frac{\pi}{4}
		\frac{\Delta m^2_{21}}{E}
		\frac{ \left( \sin 2 \eta + \frac{A_{NC}}{\Delta m^2_{21}}
			\cos 2 \eta \sin \theta \right)^2}
			{\cos 2 \eta}
		\frac{1}{\left| \frac{1}{A_{eff}}
			\frac{d A_{eff}}{dx} \right|} \right] 
		\right|_{x=x_R} \nonumber \\
& = & \left. \exp \left[- \frac{\pi}{4}
	\frac{\Delta m^2_{21}}{E}
	\frac{\sin^2 2 \theta}{\left| \frac{1}{A_{eff}} 
	\frac{d A_{eff}}{dx} \right|}
	\frac{\left(1 - \frac{|A_{NC}|}{2 \Delta m^2_{21}} 
		\cos^2 \theta \right)^2}
		{\cos^4 \theta (1 + \sin^2 \theta)} \right] 
		\right|_{x = x_R},
\end{eqnarray}
where we have used various relations in Table \ref{bonetable}.
Hence, given $\theta$ and $\phi$'s  minute effects 
on the adiabatic edge and thus the 
fitted $\Delta m^2_{21}$, if $\sin^2 2 \varphi$ fits the data for the case 
where $\theta = \phi$, the case
$\nu_{1-} \leftrightarrow \nu_{2+}$ with $\phi = 0$ will be well described by
the mixing parameter $\sin^2 2 \theta$, which is approximately related to 
$\sin^2 2 \varphi$, assuming small intergenerational mixing,
in the following manner:
\begin{eqnarray}
\label{eqb130}
\sin^2 2 \theta & \cong & \left( \frac{1}{1 - 
			\frac{|A_{NC}|}{2 \Delta m^2_{21}}} \right)^2
			\sin^2 2 \varphi 
	 \cong  \left( \frac{1}{1 - 
		\frac{N_n(x_R)}{2 N_e(x_R)}} \right)^2
		\sin^2 2 \varphi \nonumber \\
	& \cong & (1.2 \to 1.8) \sin^2 2 \varphi.
\end{eqnarray}
The same analysis applies to the case of $\nu_{1+} \leftrightarrow \nu_{2-}$.

\subsection{Case B2}

Case B2 corresponds to 
\begin{equation}
\label{eqb201}
\Delta m^2_{2+2-} \gg \Delta m^2_{21} \gg \Delta m^2_{1+1-} 
\stackrel{>}{\sim} 10^{-10} eV^2.
\end{equation}
The squared masses of the instantaneous mass eigenstates for the 
relevant solar densities are shown in
Fig.\ \ref{lss}.  As in Case B1, 
the evolution of $\nu^m_{2h}$ is virtually
density-independent and thus adiabatic due to a large $\Delta m^2_{2+2-}$.
To a very good approximation,
\begin{equation}
\label{eqb202}
\nu^m_{2h} = \nu_{2h},
\end{equation}
and
\begin{equation}
\label{eqb203}
U^m_{e2h} = U_{e2h}.
\end{equation}
The decoupling of $\nu_{2h}$ renders the $\nu_e$ survival probability 
into the form:
\begin{eqnarray}
\label{eqb204}
P^{\oplus}_{ee} (E) & = & |U_{e2h}|^4 +
		\left| \sum_{i, \, j = 1 \ell, \, 1h, \, 2 \ell}
		U^m_{ei}(x_i)
	\exp \left[- i \int^{x_f}_{x_i} {\cal H}^m_{ij} dx \right]
		U_{ej} \right|^2 \nonumber \\
	& = & |U_{e2h}|^4 + (1 - |U_{e2h}|^2)^2 P_{3 \nu MSW},
\end{eqnarray}
where
\begin{equation}
\label{eqb205}
P_{3 \nu MSW} = \left| \sum_{i, \, j = 1 \ell, \, 1h, \, 2 \ell}
\frac{U^m_{ei}}{\sqrt{U^{m2}_{e1 \ell} + U^{m2}_{e1h} + U^{m2}_{e 2 \ell}}}
\exp \left[- i  \int^{x_f}_{x_i} {\cal H}^m_{ij} dx \right]
	\frac{U_{ej}}{\sqrt{U^2_{e1 \ell} + U^2_{e1h} + U^2_{e 2 \ell}}}
	\right|^2,
\end{equation}
and $U^{m2}_{e 1 \ell} + U^{m2}_{e1h} + U^{m2}_{e 2 \ell}
= U^2_{e 1 \ell} + U^2_{e1h} + U^2_{e 2 \ell}$ by unitarity.
If we regard the decoupled $\nu^m_{2h}$ ($\sim \nu_{2h}$) state as containing
a fraction $|U^m_{e2h}|^2$ ($\sim |U_{e2h}|^2 \sim 0$) of the original $\nu_e$
population, the other $(1 - |U_{e2h}|^2 \sim 1)$ is thus distributed in the 
remaining
three states.
The splittings between these states lie, by assumption,  within the MSW range 
given in Eq.(\ref{mswmass}), forming a $3\nu$ subsystem
which undergoes, technically, two resonances $R_H$ and $R_L$.  In this
manner,
$P_{3 \nu MSW}$ is equivalent to the $\nu_e$ survival probability for a 
standard $\nu_e \leftrightarrow \nu_{\mu} \leftrightarrow \nu_{\tau}$
system with $m^2_{\tau} \gg m^2_{\mu} \simeq m^2_e$
\cite{kuo87,kuo86,3nu1,3nu2} (plus some
non-standard
features due to the presence of sterile neutrinos).  Contrary to the standard 
$3\nu$ system where
the spotlight is on $R_L$, and $R_H$ occurs at too high a density to be
relevant, our focus is on $R_H$.  (We do not consider $R_L$ because the close
encounter of the mass eigenvalues of $\nu^m_{1 \ell}$
and $\nu^m_{1h}$ in Fig. \ref{lss} represents maximal conversion of 
$\nu_e$ into
$\nu'_e$.  This happens in vacuum, necessarily adiabatically.)

Standard $3\nu$ systems have been studied 
extensively \cite{kuo87,kuo86,3nu1,3nu2}.
Following from these analyses, we parameterise the
following mixing angles:
\begin{eqnarray}
\label{eqb206}
\cos \psi_m = \frac{U^m_{e 1 \ell}}
			{\sqrt{U^{m2}_{e1 \ell} + U^{m2}_{e1h}}}, & \; &
\sin \psi_m = \frac{U^m_{e 1h}}
			{\sqrt{U^{m2}_{e1 \ell} + U^{m2}_{e1h}}}, \nonumber \\
\cos \eta_m = \sqrt{\frac{U^{m2}_{e1 \ell} + U^{m2}_{e1h}}
			{U^{m2}_{e1 \ell} + U^{m2}_{e1h} 
			+ U^{m2}_{e2 \ell}}}, & \; &
\sin \eta_m = \frac{U^m_{e2 \ell}}
			{\sqrt{U^{m2}_{e1 \ell} + U^{m2}_{e1h} 
			+ U^{m2}_{e2 \ell}}},
\end{eqnarray}
where the subscripts and superscripts $m$ denote their 
density-dependent counterparts.  The angle $\psi_m$ describes 
(approximately) the mixing
of $\nu_e$ and $\nu'_e$, and takes on a value of 
$\psi_m = \psi \cong \frac{\pi}{4}$ in vacuum.  
This mixing mode is strongly suppressed at high 
densities because of the small vacuum splitting between $\nu_{1 \ell}$ and 
$\nu_{1h}$ (so that $\psi_m \rightarrow \frac{\pi}{2}$ as
$\rho \rightarrow \infty$).  Consequently, the $\nu_1$ state that takes part 
in the intergenerational MSW resonance $R_H$ at density 
$\rho \sim \Delta m^2_{21}$ resembles neither $\nu_{1 \ell}$ nor $\nu_{1h}$
but is, asymptotically, some approximately maximal linear combinations of the
two states, which we denote as $\nu_{1x}$.  Its orthogonal state, $\nu_{1y}$,
is thus the asymptotic form of $\nu^m_{1 \ell}$ for the relevant solar 
densities, that is,
\begin{equation}
\label{eqb207}
\left( \begin{array}{c}
	\nu_{1x} \\
	\nu_{1y} \\
	\nu_{2 \ell} \\
	\nu_{2h} \\
	\end{array} \right)
= {\cal O}(\psi) \left( \begin{array}{c}
		\nu_{1 \ell} \\
		\nu_{1h} \\
		\nu_{2 \ell} \\	
		\nu_{2h} \\	
		\end{array} \right)
= \left( \begin{array}{cccc}
		\cos \psi & \sin \psi & 0 & 0 \\
		- \sin \psi & \cos \psi & 0 & 0 \\
		0 & 0 & 1 & 0 \\
		0 & 0 & 0 & 1 \\
		\end{array} \right)
		\left( \begin{array}{c}
		\nu_{1 \ell} \\
		\nu_{1h} \\
		\nu_{2 \ell} \\	
		\nu_{2h} \\	
		\end{array} \right).
\end{equation}
Note that for small intergenerational mixing, $\nu_{1x} \simeq \nu_e$ and
$\nu_{1y} \simeq \nu'_e$ by Eqs. (\ref{eqb206}) and (\ref{eqb207}).
At density $0 \ll \rho \ll \rho_{R_H}$, we may
treat the $4\nu$ system as consisting of two parity eigenstates
$\nu_{2 \ell}$ and $\nu_{2h}$, and two pseudo-mass eigenstates $\nu_{1x}$ and
$\nu_{1y}$ of indefinite parity.

The angle $\eta$ describes the {\it apparent} mixing of $\nu_{1x}$ and 
$\nu_{2 \ell}$ in vacuum, which is minimal as inferred from 
Eqs.(\ref{4numatrix}) and (\ref{eqb206}) assuming small intergenerational 
mixing.  However, in the proximity of $R_H$ (at $\rho \sim \Delta m^2_{21}$),
while $\nu_{1 y}$ and $\nu_{2h}$ propagate virtually density-independently, 
matter effects rotate $\eta_m$ from its vacuum value $\eta$ through 
$\frac{\pi}{4}$ to $\frac{\pi}{2}$, and thereby modify the evolution of
$\nu_{1x}$ and $\nu_{2 \ell}$ dramatically.  If we regard $\nu_{1x}$ and 
$\nu_{2 \ell}$ as forming two orthogonal pseudo-weak eigenstates $\nu_a$ and 
$\nu_b$ that convert resonantly into each other at $R_H$, i.e.,
\begin{eqnarray}
\label{eqb208}
\left( \begin{array}{c}
	\nu_a \\
	\nu_{1y} \\
	\nu_b \\
	\nu_{2h} \\
	\end{array} \right)
& = & {\cal S}(\eta) \left( \begin{array} {c}
			\nu_{1x} \\
			\nu_{1y} \\	
			\nu_{2 \ell} \\
			\nu_{2h} \\
			\end{array} \right)
= \left( \begin{array}{cccc}
	\cos \eta & 0 & \sin \eta & 0 \\
	0 & 1 & 0 & 0 \\
	- \sin \eta & 0 & \cos \eta & 0 \\
	0 & 0 & 0 & 1 \\
	\end{array} \right)
	\left( \begin{array} {c}
			\nu_{1x} \\
			\nu_{1y} \\	
			\nu_{2 \ell} \\
			\nu_{2h} \\
			\end{array} \right) \nonumber \\
& = & {\cal S}(\eta) {\cal O}(\psi) \left( \begin{array} {c}
					\nu_{1 \ell} \\
					\nu_{1h} \\	
					\nu_{2 \ell} \\
					\nu_{2h} \\
					\end{array} \right)
= \left( \begin{array}{cccc}
	\cos \eta \cos \psi & \cos \eta \sin \psi & \sin \eta & 0 \\
	- \sin \psi & \cos \psi & 0 & 0 \\
	- \sin \eta \cos \psi & - \sin \eta \sin \psi & \cos \eta & 0 \\
	0 & 0 & 0 & 1 \\
	\end{array} \right)
		\left( \begin{array} {c}
				\nu_{1 \ell} \\
				\nu_{1h} \\	
				\nu_{2 \ell} \\
				\nu_{2h} \\
				\end{array} \right),
\end{eqnarray}
we may follow a similar procedure to Eqs.(\ref{eqb110}) and (\ref{eqb111})
and approximate $U^m$ as
\begin{equation}
\label{eqb209}
U^m = U {\cal O}^{-1}(\psi) {\cal S}^{-1}(\eta)
	{\cal S}(\eta_m) {\cal O}(\psi_m),
\end{equation}
such that the effective Hamiltonian that governs the evolution of $\nu_a$
and $\nu_b$ in the vicinity of $R_H$ is given by the $1-3$ sector of
\begin{eqnarray}
\label{eqb210}
{\cal H}_{eff} & = & \frac{1}{2E} [ {\cal S}(\eta) {\cal O}(\psi) {\cal M}
				{\cal O}^{-1}(\psi) {\cal S}^{-1}(\eta) + 
		{\cal S}(\eta) {\cal O}(\psi) U^{-1} {\cal H}_{int}U
			{\cal O}^{-1}(\psi) {\cal S}^{-1}(\eta)] \nonumber \\
& \stackrel{1-3 sector}{\rightarrow} & \frac{1}{4}
	\left( \begin{array}{cc}
		- \Delta m^2_{eff} \cos 2 \eta + A_{eff}  
		& \Delta m^2_{eff} \sin 2 \eta + A_{ind} \\
		\Delta m^2_{eff} \sin 2 \eta + A_{ind} 
		& \Delta m^2_{eff} \cos 2 \eta - A_{eff} \\
		\end{array} \right) +  constant,
\end{eqnarray}
where
\begin{equation}
\label{eqb211}
\Delta m^2_{eff} \equiv m^2_{2 \ell} 
		- \frac{1}{2}(m^2_{1 \ell} + m^2_{1h} 
					- \Delta m^2_{1+1-} \cos 2 \psi)
		\cong m^2_{2 \ell} - \bar{m}^2_1.
\end{equation}
The term $\Delta m^2_{eff}$ represents the effective vacuum squared mass 
difference
responsible for the resonance, and $\bar{m}^2_1$ is the averaged squared mass
of the $\nu_1$ states, $\bar{m}^2_1 = \frac{1}{2}(m^2_{1 \ell} + m^2_{1h})$.  
The magnitudes of $\Delta m^2_{eff}$ and $\Delta m^2_{21}$, where
$\Delta m^2_{21} = m^2_{2 \ell} - m^2_{1h}$, are virtually identical.
Henceforth, we shall replace $\Delta m^2_{eff}$ with $\Delta m^2_{21}$
whenever the former is encountered for convenience.  The quantities
$A_{eff}$ and $A_{ind}$ are given by \begin{eqnarray}
\label{eqb212a}
A_{eff} & = & A_{CC} (U^2_{e1 \ell} + U^2_{e1h} + U^2_{e2 \ell}) \nonumber \\
& + & A_{NC} \{ \cos 2 \eta [(\cos \psi U_{e 1 \ell} 
		+ \sin \psi U_{e1h})^2 - U^2_{e 2 \ell}
	+ (\cos \psi U_{\mu 1 \ell} + \sin \psi U_{\mu 1h})^2 - 
		U^2_{\mu 2 \ell}] \nonumber \\
& + & 2 \sin 2 \eta [U_{e2 \ell}(\cos \psi U_{e1 \ell} 
		+ \sin \psi U_{e1h})
	+ U_{\mu2 \ell} (\cos \psi U_{\mu1 \ell} + 
		\sin \psi U_{\mu 1h})] \},
\end{eqnarray}
and
\begin{eqnarray}
\label{eqb212b}
A_{ind} & = & A_{NC} \{ 2 \cos 2 \eta [U_{e2 \ell}(\cos \psi U_{e1 \ell}
			+ \sin \psi U_{e1h})
		+ U_{\mu 2 \ell} (\cos \psi U_{\mu 1 \ell} 
		+ \sin \psi U_{\mu 1h})] \nonumber \\
	& - & \sin 2 \eta [(\cos \psi U_{e1 \ell} + \sin \psi U_{e1h})^2
			- U^2_{e2 \ell}
	+ (\cos \psi U_{\mu 1 \ell} + \sin \psi U_{\mu 1h})^2
		- U^2_{\mu 2 \ell}] \},
\end{eqnarray}
respectively.  Table \ref{btwotable} shows 
$A_{eff}$ and $A_{ind}$ evaluated for various 
combinations of $\nu_{1 \ell}, \, \nu_{1h}$ and $\nu_{2 \ell}$. 

Taking the well-established $3\nu$ survival probability 
from Ref.\cite{kuo87} and 
setting the level-crossing probability at $R_L$ to zero, we may
immediately write down $P_{3 \nu MSW}$ as
\begin{eqnarray}
\label{eqb213}
P_{3 \nu MSW} & = & \sin^2 \eta_m (x_i) \sin^2 \eta 
		+ \cos^2 \eta_m (x_i) \cos^2 \eta
		[\sin^2 \psi_m (x_i) \sin^2 \psi 
		+ \cos^2 \psi_m (x_i) \cos^2 \psi]
		\nonumber \\
	& - & P_{R_H} [\sin^2 \eta_m (x_i) 
		- \cos^2 \eta_m (x_i) \sin^2 \psi_m (x_i)]
		(\sin^2 \eta - \cos^2 \eta \sin^2 \psi),
\end{eqnarray}
where $P_{R_H}$ is the level-crossing probability at resonance $R_H$
calculated from ${\cal H}_{eff}$ in Eq.(\ref{eqb210}).  
Furthermore, because of the 
strong suppression
of $\nu_e \leftrightarrow \nu'_e$ oscillations in most parts of the sun,
the corresponding matter mixing angle $\psi_m$ is close to 
$\frac{\pi}{2}$, thereby reducing
Eq.(\ref{eqb213}) to
\begin{eqnarray}
\label{eqb214}
P_{3 \nu MSW} & = & \sin^2 \eta_m (x_i) \sin^2 \eta 
			+ \cos^2 \eta_m (x_i) \cos^2 \eta \sin^2 \psi 
			\nonumber \\
		&-  & P_{R_H} [\sin^2 \eta_m (x_i)- \cos^2 \eta_m (x_i)]
			(\sin^2 \eta - \cos^2 \eta \sin^2 \psi).
\end{eqnarray}
Equation (\ref{eqb214}) does not present itself in the most illuminating for 
the purpose of comparison.  In the first
instance, it does not, superficially, resemble the familiar expression for the 
standard $2 \nu$ MSW transition probability $P_{2 \nu MSW}$ 
in Eq.(\ref{eqi}).  However, putting it into context, the overall $\nu_e$ 
survival probability $P^{\oplus}_{ee}(E)$ for Case B2 in Eq.(\ref{eqb204}) 
can be recast into
\begin{equation}
\label{eqb215}
P^{\oplus}_{ee} (E) 
= |U_{e2h}|^4 + (1 - |U_{e2h}|^2) (1 - |U_{e1 \ell}|^2 - |U_{e2h}|^2)
		\tilde{P}_{2 \nu MSW},
\end{equation}
where
\begin{equation}
\label{eqb216}
\tilde{P}_{2 \nu MSW} = \frac{1}{2} + (\frac{1}{2} - P_{R_H})
			\cos 2 \eta_m (x_i) \cos 2 \omega,
\end{equation}
and
\begin{equation}
\label{eqb217}
\cos \omega = \frac{U_{e 1h}}{\sqrt{U^2_{e1h} + U^2_{e 2 \ell}}}, \;
\sin \omega = \frac{U_{e 2 \ell}}{\sqrt{U^2_{e1h} + U^2_{e 2 \ell}}}.
\end{equation}
In the limit of small intergenerational mixing, 
$\tilde{P}_{2 \nu MSW} \rightarrow P_{2 \nu MSW}$ in such a way that 
Eq.(\ref{eqb215}) is well approximated by
\begin{equation}
\label{eqb218}
P^{\oplus}_{ee}(E) \approx \frac{1}{2} P_{2 \nu MSW},
\end{equation}
where $P_{2 \nu MSW}$ is evaluated for $\Delta m^2_{21}$ and 
$\eta$ in a medium of effective density 
\begin{equation}
\label{eqb219}
A_{eff} \approx A_{CC} + \frac{1}{2} A_{NC},
\end{equation}
as inferred from Table \ref{btwotable}, 
with a level-crossing probability $P_{R_H}$
governed by ${\cal H}_{eff}$ in Eq.(\ref{eqb210}).
Given the respective standard 
$\nu_e \leftrightarrow \nu_{\mu, \, \tau}$ and $\nu_e \leftrightarrow \nu_s$
effective densities:
\begin{eqnarray}
\label{eqb220}
A_{eff}(\nu_e \leftrightarrow \nu_{\mu, \, \tau}) & = & A_{CC}, \nonumber \\
A_{eff}(\nu_e \leftrightarrow \nu_s) & = & A_{CC} + A_{NC},
\end{eqnarray}
Eq.(\ref{eqb219}) immediately puts the adiabatic edge of the 
MSW transition between that
of the two standard cases for the same oscillation parameters
as shown in Fig. \ref{lssprob}. 

On the non-adiabatic aspect of $P_{2 \nu MSW}$, we learn from
Table \ref{btwotable} that the mixing angle $\eta$ 
and the off-diagonal term $A_{ind}$
(both of which appear in the level-crossing probability)  
do not distinguish between the exact identities of 
$\nu_{1 \ell}$ and $\nu_{1h}$, or equivalently, the sign of 
$\Delta m^2_{1+1-}$.  
(They do, however, depend on the identity of 
$\nu_{2 \ell}$ --- a matter of interchanging 
$\theta$ and $\phi$.)  This is because
the enhanced {\it apparent} mixing mode at $R_H$
is that of $\nu_{1x} \leftrightarrow \nu_{2 \ell}$.  The state $\nu_{1x}$, 
in turn, is
a fixed, $\Delta m^2_{1+1-}$ independent admixture of $\nu_{1+}$ and 
$\nu_{1-}$, such that $\Delta m^2_{1+1-}$ does not enter the scene so long
as Eq.(\ref{eqb201}) is satisfied.  
The matter-induced mixing term $A_{ind}$ varies with $\theta$ and $\phi$.
Taking the case of $\nu_{1x} \leftrightarrow \nu_{2-}$, we observe that 
$A_{ind}$ vanishes exactly when $\theta = 0$.  For $\theta = \phi$, we have
$A_{ind} = - \frac{A_{NC}}{2} \sin 2 \eta = \frac{|A_{NC}|}{2} \sin 2 \eta$
so that the ratio of matter-induced to vacuum mixing evaluated at resonance
is given by
\begin{equation}
\label{eqb221}
\Lambda = \frac{|A_{NC}|}{2 \Delta m^2_{21}} 
= \frac{\cos 2 \eta}{4 \left( \frac{N_e(x_R)}{N_n(x_R)} \right) -1}
\approx 0.04 \to 0.14.
\end{equation}
The contribution of matter-induced mixing 
(which is additive here by definition)
is therefore relatively small.  Given an effective density in 
Eq.(\ref{eqb219}) that is almost independent of $\theta$ and $\phi$ 
(provided they are small), approximately the same 
$\Delta m^2_{21}$ will provide a fit to the experimental data for any
combination of $\theta$ and $\phi$.  It then follows from 
Eq.(\ref{eqb221}) that for the $\theta = \phi$ case, $\sin^2 2 \eta$  
lies in the range
\begin{equation}
\label{eqb222}
\sin^2 2 \eta \approx (0.8 \to 0.9) \sin^2 2 \varphi,
\end{equation}
where $\sin^2 2 \varphi$ is the fitted mixing parameter for the case where 
no matter-induced mixing is present (i.e., when $\theta = 0$).  
At the other extreme, Table \ref{btwotable} shows that for 
$\nu_{1x} \leftrightarrow \nu_{2-}$, matter-induced 
mixing takes full control if $\phi = 0$.  This mixing, however, is negligible.
Following from Eqs.(\ref{eqb120}) to (\ref{eqb122}), 
the quantity $\kappa^2(x)$
that appears in the level-crossing probability is now proportional to 
$A^2_{ind}$, i.e.,
\begin{eqnarray}
\label{eqb223}
\left( |A_{NC}| \cos 2 \eta \cos \alpha \sin \theta \right)^2
& = & (\Delta m^2_{21})^2 \sin^2 2 \theta 
	\left( \frac{|A_{NC}|}{2 \Delta m^2_{21}} \right)^2
	\frac{1}{\cos^2 \theta + 1} \nonumber \\
& \approx & (8 \times 10^{-4} \to 10^{-2}) 
		(\Delta m^2_{21})^2 \sin^2 2 \theta.
\end{eqnarray}
Thus, given that $\sin^2 2 \theta$ is sufficiently small, the level-crossing
probability for $\phi = 0$ is almost one.  
[If the resonance is sufficiently close to the centre
of the sun where $\frac{N_e}{N_n} \sim 2$, matter-induced mixing may show 
itself by contributing to the level-crossing probability an equivalent
of $\sin^2 2 \theta \times 10^{-2}$ by Eq.(\ref{eqb223}).]

We may carry out the same analysis for $\nu_{1x} \leftrightarrow \nu_{2+}$,
which involves little more than interchanging $\theta$ and $\phi$.

\subsection{Case C}

Case C comprises the following parameters:
\begin{equation}
\label{eqc01}
\Delta m^2_{2+2-} \gg \Delta m^2_{21} \gg \Delta m^2_{1+1-},
\; \Delta m^2_{1+1-} \stackrel{<}{\sim} 10^{-11} eV^2.
\end{equation}
The mathematics that describes the resonant conversion of $\nu_e$ in the 
interior of the sun for this case is identical to that for Case B2 where the
resonance $R_H$ is one of enhanced $\nu_{1x} \leftrightarrow \nu_{2 \ell}$
mixing.  
Indeed, the two systems are physically identical.  However, the suppression of 
$\nu_e \leftrightarrow  \nu'_e$ oscillations is no longer solely a matter 
effect but is extended to the vacuum, leading to a vastly different 
phenomenology.  Vacuum $\nu_e \leftrightarrow \nu'_e$ oscillations do not 
happen because of the extremely small $\Delta m^2_{1+1-}$ which in turn  
corresponds to an oscillation length much longer than an astronomical unit.
We may therefore treat the problem as though the lower resonance $R_L$ 
(responsible for $\nu_e \leftrightarrow \nu'_e$) in Fig. \ref{lss} 
is completely
absent.  The $\nu_e$ survival probability is then given by Eq.(\ref{eqb204})
with $P_{3 \nu MSW}$ replaced with $P_{2 \nu MSW}$, i.e.,
\begin{equation}
\label{eqc02}
P^{\oplus}_{ee}(E) = |U_{e2h}|^4 + (1 - |U_{e2h}|^2)^2 P_{2 \nu MSW}.
\end{equation}
The MSW transition probability $P_{2 \nu MSW}$ is that in Eq.(\ref{eqi}) 
evaluated for the effective Hamiltonian of Case B2 in Eq.(\ref{eqb210}) with
a squared mass difference $\Delta m^2_{eff} \cong \Delta m^2_{21}$ and
mixing angle $\eta$ defined in Eqs.(\ref{eqb211}) and 
(\ref{eqb206}) respectively.  
Assuming small 
intergenerational mixing, Eq.(\ref{eqc02}) reduces to
\begin{equation}
\label{eqc03}
P^{\oplus}_{ee} (E) \approx P_{2 \nu MSW},
\end{equation}
where $P_{2 \nu MSW}$ contains non-standard features as described earlier
for Case B2.  Equation (\ref{eqc03}) is represented graphically in 
Fig. \ref{novac}.

\section{Solutions to the solar neutrino problem}
\label{sec4}

In this section, we locate the regions of parameter space that
will give rise to the observed solar neutrino depletion for Cases B1, B2 and C
in an approximate way.  
Currently available experimental data suggest a significant depletion of the
mid-energy neutrinos, hinting at a pre-defined shape for the $\nu_e$ 
survival probability.  By going to regions of parameter space in which 
matter-induced mixing is small, the 
approximate $\nu_e$ survival probabilities in Eqs.(\ref{eqb119a}), 
(\ref{eqb218}) and (\ref{eqc03}) for Cases 
B1, B2 and C respectively are very simply related to the standard two-flavour 
$P_{2 \nu MSW}$.  Thus, by comparison with the standard solutions, we
may gain a rough feeling 
for the necessary oscillations parameters for each case without performing an
{\it ab initio} fit to the experimental data.   

Flux-independent data from Kamiokande and SuperKamiokande such as spectral 
distortion and day-night asymmetry provide yet another means to identify the 
allowed oscillation parameters \cite{hl,flm3,flm2}.  
Although the day-night effect is beyond the scope of this paper, we will be
able to comment on the expected spectral distortion in Cases B1, B2 and C.

\subsection{Case B1}

Let us reiterate that the oscillation parameters for Case B1 are constrained
by Eqs.(\ref{eqb101}) and (\ref{eqb102}).  For comparison, 
it is useful to define the
ratio of the event rate with oscillations to that with no oscillations as
\begin{equation}
\label{eqsolb101}
\Omega = \frac{\int^{\infty}_{E_0} \overline{P}^{\oplus}_{ee}(E) 
	\phi^{\oplus}_0 (e, E) \sigma(E)dE}
	{\int^{\infty}_{E_0}  \phi^{\oplus}_0 (e, E) \sigma(E) dE}, 
\end{equation}
where $\phi^{\oplus}_0 (e, E)$ is the no-oscillation $\nu_e$ flux,
$\sigma (E)$ is the detection cross-section, $E_0$ is the experimental
energy threshold, and $\overline{P}^{\oplus}_{ee} (E)$ is the $\nu_e$
survival probability averaged over production positions.
By Eq.(\ref{eqb119a}), the energy-dependence of the $\nu_e$ survival
probability for this case is contained entirely in the term $P_{2 \nu MSW}$.  
In the small intergenerational mixing limit, we further approximate
the adiabatic edge of $P_{2 \nu MSW}$ as a step function 
$\theta(E^{4 \nu}_{A} - E)$, where $E^{4 \nu}_A$ is $E_A$ defined in 
Eq.(\ref{eqb119}), such that
\begin{equation}
\label{eqsolb102}
P_{2 \nu MSW} |_{4 \nu} \cong \theta(E^{4 \nu}_A - E) + P^{4 \nu}_R.
\end{equation} 
For comparison purposes, we also write down an analogous expression for a
$2\nu$ system:
\begin{equation}
\label{eqsolb103}
P_{2 \nu MSW} |_{2 \nu} \cong \theta(E^{2 \nu}_A - E) + P^{2 \nu}_R,
\end{equation}
where 
\begin{equation}
\label{eqsolb103a}
E^{2 \nu}_A = \frac{\Delta m^2_{21} \cos 2 \eta}{2 \sqrt{2} G_F N_e(x_i)}.
\end{equation}
Equation (\ref{eqsolb103a}) comes from the resonance condition, 
$A_{eff} = A_{CC} = \Delta m^2_{21} \cos 2 \eta$, for a standard 
$\nu_e \leftrightarrow \nu_{\mu, \, \tau}$ 
system, and a comparison with Eq.(\ref{eqb119}) immediately leads us to
\begin{equation}
\label{eqsolb103b}
2 E^{2 \nu}_A \cong  E^{4 \nu}_A.
\end{equation}
The scale heights 
$\left| \frac{1}{A_{eff}} \frac{dA_{eff}}{dt} \right|$ that
appear in the level-crossing probabilities $P^{4 \nu}_R$ and $P^{2 \nu}_{R}$
do not differ much due to the almost exponential 
solar density profile \cite{bp95}.  Thus, if we ignore matter-induced 
mixing by setting 
$\theta = \phi$ according to Table \ref{bonetable}, 
we may make the approximation:
\begin{equation}
\label{eqsolb103c}  
P^{4 \nu}_R \cong P^{2 \nu}_R, 
\end{equation}
such that, by Eqs.(\ref{eqb119a}), (\ref{eqsolb101} -- \ref{eqsolb103}) and
(\ref{eqsolb103b}),
\begin{equation}
\label{eqsolb104}
\Omega_{4 \nu}(\Delta m^2, \sin^2 2 \eta) 
\simeq \frac{1}{4} + 
\frac{1}{4} \Omega_{2 \nu} (2 \Delta m^2, \frac{1}{2} \sin^2 2 \eta).
\end{equation}
Here, $\Omega_{4 \nu}(\Delta m^2, \sin^2 2 \eta)$ denotes the value of 
$\Omega$ in our 
$4\nu$ scheme evaluated for $\Delta m^2$ and $\sin^2 2 \eta$, and 
$\Omega_{2 \nu}(2 \Delta m^2, \frac{1}{2} \sin^2 2 \eta)$ is the ratio 
$\Omega$ in the 
standard $\nu_e \leftrightarrow \nu_{\mu, \, \tau}$ scheme evaluated at
twice the squared mass 
difference and half the mixing.  The rescaling of $\Delta m^2$ and 
$\sin^2 2 \eta$ in the latter is dictated by Eq.(\ref{eqsolb103c}),
such that the parameter $\Delta m^2 \sin^2 2 \eta$ that is fed into 
$P^{4 \nu}_{R}$ and $P^{2 \nu}_r$ respectively agree.  
Thus, in the extreme case of  
$\Delta m^2_{1+1-} \sim 10^{-3} eV^2$,
a minimum of $\frac{1}{4}$ of the original neutrino flux must be detected, 
while the maximum detectable flux is $\frac{1}{2}$ as a direct consequence of 
maximal vacuum $\nu_e \leftrightarrow \nu'_e$ oscillations.  The latter 
corresponds to the absence of MSW transitions, or equivalently, to 
$\Delta m^2_{21}$ and $\eta$ residing in regions of parameter space outside 
of that quoted in Eqs.(\ref{mswmass}) and (\ref{mswmixing}) respectively.

Equation (\ref{eqsolb104}) allows us to virtually fit the experimental data 
by the use 
of existing theoretical $\nu_e \leftrightarrow \nu_{\mu, \, \tau}$ 
predictions for the various 
experiments in which smearing over production point and other energy 
dependence are already accounted for.  Given a set of standard 
$2\nu$ MSW contours, we can pick out the necessary oscillation 
parameters for this case graphically by identifying each 
$\Omega_{2 \nu}$ contour with a $\Omega_{4 \nu}$ by Eq.(\ref{eqsolb104}), and 
adjusting the 
$\Delta m^2$ and $\sin ^2 2 \eta$ scales accordingly.  The area enclosed by
the dot-dash line in
Fig. \ref{myfit} represents the region of parameter space in 
which the $2\sigma$
bands (including both experimental and theoretical errors) of all five 
experiments overlap \cite{flm}.  Numerically, this region
is defined by
\begin{equation}
2 \times 10^{-6} \stackrel{<}{\sim} \Delta m^2_{21}/eV^2 \stackrel{<}{\sim}
5 \times 10^{-5},
\end{equation}
and
\begin{equation}
8 \times 10^{-4} \stackrel{<}{\sim} \sin^2 2 \eta \stackrel{<}{\sim} 0.1.
\end{equation}
[Beyond $\sin^2 2 \eta \sim 0.1$, Eqs.(\ref{eqb119a}) and 
(\ref{eqsolb104}) become invalid as intergenerational vacuum oscillations 
increase in amplitude.]  
Note that our fitting procedure is approximate, but should nevertheless 
yield a solid indication of the $2\sigma$ allowed parameter space. 
We also find that the $1\sigma$ bands do not overlap in this 
case, suggesting that the solution is not acceptable at below $68 \%$ C. L..
Figure \ref{lslfit} shows the $\nu_e$ survival probability 
for several representative sets of oscillation parameters within
the allowed
region for Case B1.

A qualitative discussion of this virtual fit follows.  Firstly, from 
Eq. (\ref{SKrange}), 
to Table \ref{expresults}, 
large uncertainties in the Boron flux means
that Kamiokande and SuperKamiokande's $2\sigma$ bands 
necessarily span the entire 
region of MSW parameter space for this case.  
Secondly, the central values of the GALLEX 
and SAGE results are somewhat higher than the maximum detectable ratio of 
$\frac{1}{2}$ as predicted by the EPM.  Hence, the lower bound on 
$\Delta m^2_{21}$ is set by the Gallium experiments, corresponding to 
the maximum amount of low- and mid-energy neutrinos that 
can be killed resonantly 
within limits.        
 
The central value of the Homestake result is $\sim \frac{1}{4}$ relative to 
SSM prediction, while at plus $2\sigma$, the ratio of
measured to no-oscillation event rates does 
not quite reach the maximum of $\frac{1}{2}$.  
Thus, the upper bound on the squared 
mass difference $\Delta m^2_{21}$ and the lower bound on 
$\Delta m^2_{21} \sin^2 2 \eta$ are 
determined by Homestake, representing the possible suppression patterns that
the Boron spectrum may receive.  

Note that we have arrived at these assuming $\theta = \phi$.  A rough 
indication of the necessary regions of parameter space for cases of unequal 
$\theta$ and $\phi$ can be obtained based on the analyses in Sec. \ref{sec3}.
In short, unequal $\theta$ and $\phi$ will give more breadth to the allowed
region.

\subsection{Case B2}

In Case B2, the parameters are constrained by Eq.(\ref{eqb201}).  The 
$\nu_e$ survival probability for this case in Eq.(\ref{eqb215}) consists of
negligible constant terms due to averaged vacuum 
oscillations.  
This is a direct consequence of 
the strong suppression of $\nu_e \leftrightarrow \nu'_e$ oscillations in 
matter such that almost all of the original $\nu_e$'s produced at 
$\rho \gg \rho_R$ will pass through a resonance.
In this case, 
$P^{\oplus}_{ee}(E)$ plunges down to as low as $\sim 0$ 
immediately to the high-energy end  of the adiabatic edge according to 
Eq.(\ref{eqb218}) and Fig. \ref{lssprob}.  The correspondingly deeper pit 
in $P_{2 \nu MSW}$ implies 
that the adiabatic edge needs to occur at an even higher energy in order to 
maximise the number of low- and mid-energy neutrinos to be detected.  
This is attainable by choosing an even higher squared mass difference.

The naturally smaller-than-standard effective density in Eq.(\ref{eqb219}) 
places the adiabatic 
edge of Case B2 at a somewhat higher energy.  The exact
location of the edge 
relative to a standard edge, however, cannot be simply quantified 
since $N_e(x)$ and $N_n(x)$ do not exactly track
each other
in the sun.  However, expressing $A_{eff}$ of Eq.(\ref{eqb219}) in terms 
of $A_{CC}$  
for the relevant solar densities,  
\begin{equation}
\label{eqsolb201}
A_{eff} \approx (\frac{7}{8} \to \frac{23}{24}) A_{CC},
\end{equation}
the shift of the adiabatic edge relative to the standard location is
negligible.  On the non-adiabatic side, if we neglect matter-induced
mixing (by setting $\theta = 0$ for $\nu_{1x} \leftrightarrow \nu_{2-}$, or
$\phi = 0$ for $\nu_{1x} \leftrightarrow \nu_{2+}$), $P^{4 \nu}_R$
evaluated for this case will be approximately the same as $P^{2 \nu}_R$ for 
a standard $2\nu$ system with the same $\Delta m^2$
and $\eta$.  Hence, $P_{2 \nu MSW}|_{4 \nu} \approx P_{2 \nu MSW}|_{2 \nu}$
and
\begin{equation}
\label{eqsolb202}
\Omega_{4 \nu} (\Delta m^2, \sin^2 2 \eta)
\approx \frac{1}{2} \Omega_{2 \nu}(\Delta m^2, \sin^2 2 \eta),
\end{equation}
by Eqs.(\ref{eqb218}) and (\ref{eqsolb101}), and the symbols carry 
the same definitions as before.  Thus, utilising
established $2 \nu$  MSW transition probability contours, 
we obtain the allowed oscillation parameters for 
Case B2 by Eq.(\ref{eqsolb202}) in a region defined by
\begin{equation}
10^{-5} \stackrel{<}{\sim} \Delta m^2_{21}/eV^2 
\stackrel{<}{\sim} 10^{-4},
\end{equation}
and
\begin{equation}
10^{-4} \stackrel{<}{\sim} \sin^2 2 \eta \stackrel{<}{\sim} 4 \times 10^{-3},
\end{equation} 
including both experimental and theoretical errors at $2\sigma$ as shown in 
Fig. \ref{myfit}.    This is a considerably smaller region
than that for Case B1.  In particular, the Kamiokande and SuperKamiokande
results now place an
upper bound on the allowed $\sin^2 2 \eta$ such that we do not wipe out
too many Boron neutrinos by the MSW mechanism in the 
non-adiabatic branch.     
The $\nu_e$ survival probability
for this case for several representative sets of oscillation parameters
within the allowed region is shown in Fig. \ref{lssfit}.

\subsection{Case C}

Case C is described by the parameters given in Eq.(\ref{eqc01}).
In view of Fig. \ref{novac}, the similarity between the energy-dependences
of this case and of the standard $2\nu$ (both 
$\nu_e \leftrightarrow \nu_{\mu, \, \tau}$ and $\nu_e \leftrightarrow \nu_s$)
scenarios shows that due to  
the absence of vacuum $\nu_e \leftrightarrow \nu'_e$ oscillations, 
the necessary oscillation parameters for Case C will lie  in between those 
of the two standard cases (see Table \ref{summary}).  
The solution at $95 \%$ C. L. is shown in Fig. \ref{myfit}.

\section{Prospects for future experiments}
\label{sec5}

We now briefly discuss the implications of the various solutions on 
experimental observables.  In particular, we shall look at the Boron energy 
spectrum observed by the high-energy scattering experiments, the 
Beryllium line that will be detected by BOREXINO, and the charged to neutral
current event rates to be measured by SNO.

\subsection{Boron spectral distortion}

Qualitatively, the underlying maximal $\nu_e \leftrightarrow \nu'_e$ vacuum 
oscillations in Cases B1 and B2 will lead to recoil electron energy spectra 
for SuperKamiokande and SNO respectively that are almost flat with respect 
to SSM predictions.  
This can be seen by comparing the various survival probabilities 
in Figs. \ref{lslfit} and \ref{lssfit}. For Case B1 (Fig.\ \ref{lslfit}),
the energy-independent $\frac{1}{4}$ that provides a lower limit to the
flux depletion considerably softens the energy-dependence of the non-adiabatic
branch of the $\nu_e$ survival probability. For Case B2 (Fig.\ \ref{lssfit}),
the slope of the non-adiabatic branch is scaled down by a factor of two,
relative to the standard MSW solutions, because of Eq.(\ref{eqb218}).   
For Case A, complete energy-dependence means that spectral distortion is 
absent as in the no-oscillation case.
Precise 
predictions for the amount of deformation in energy-dependent cases 
relative to  
standard expectations cannot be obtained without performing an {\it ab initio}
numerical fit, 
because of the vastly different oscillation parameters 
involved.  However, it may 
be said for certain that the spectral distortions in Cases B1 and B2 are 
significantly weaker than those predicted by all other minimal 
$\nu_e \leftrightarrow \nu_{\mu, \, \tau}$
 and/or 
$\nu_e \leftrightarrow \nu_s$ schemes currently in the market 
\cite{distortion}, and are somewhat stronger than the standard large 
mixing angle (LMA) scheme.

Based on $\frac{\Delta T}{T}$, 
the deviation of the averaged measured electron
kinetic energy from its standard value,  
SuperKamiokande's flux-independent 
data to date do not distinguish between the standard $2\nu$
small mixing angle, 
the large mixing angle and the 
no-oscillation solutions within $1\sigma$ \cite{flm2}. From 
Figs. \ref{lslfit} and \ref{lssfit}, we expect the quantities 
$\frac{\Delta T}{T}$ resulting from Cases B1 and B2 respectively to 
take on some intermediate values, compared with the standard 
SMA and LMA solutions.  In this respect, Cases B1 and B2 
are consistent with spectral data to date.  Case A is also acceptable.      
(A small reduction in the allowed regions for Cases B1 and B2 displayed
in Fig.\ \ref{myfit} may result from a rigorous consideration of
existing spectral distortion data. If this were to occur, its cause would
be the encroachment of the adiabatic edge above the energy threshold for
SuperKamiokande, which is presently $\sim 6.5\ MeV$. An inspection of Figs.\
\ref{lslfit} and \ref{lssfit}, however, reveals that the adiabatic edges 
occur at less than $6.5\ MeV$ in our admittedly approximate fits.)

Observational effects associated with Case C are similar to those studied in 
Ref.\cite{liu}.  
This case will give a recoil electron energy spectrum that is 
similar to that predicted by the standard $2\nu$ cases.  
Currently available flux-dependent and flux-independent data 
do not distinguish between this case and 
the standard $\nu_e \leftrightarrow \nu_{\mu, \, \tau}$ 
and $\nu_e \leftrightarrow \nu_s$ SMA scenarios.  However, the 
SMA solution is preferred by SuperKamiokande over the LMA 
solution based on spectral data analyses \cite{flm2}.  To this end, Case C 
looks promising.

\subsection{The Beryllium line}

With the exception of Case C which predicts an energy-dependence that is 
similar to the standard $\nu_e \leftrightarrow \nu_{\mu, \, \tau}$ and 
$\nu_e \leftrightarrow \nu_s$ scenarios, maximal vacuum 
$\nu_e \leftrightarrow \nu'_e$ oscillations lead to a Beryllium 
line that must be detected at $\sim \frac{1}{4}$ to  
$\sim \frac{1}{2}$ the no-oscillation rate by BOREXINO.  These deductions come 
from an inspection of Figs. \ref{lslfit} and \ref{lssfit}.  In particular, 
if Case B2 is valid, the Beryllium flux will be almost exactly halved, 
independent of the intergenerational oscillation parameters (provided they 
lie within the allowed region shown in Fig. \ref{myfit}).  
In this respect, Cases B1 and B2 are clearly distinguishable from 
the standard $2 \nu$ SMA schemes.  Needless to say, Case A being 
energy-independent will exactly halve the Beryllium flux. 

\subsection{Charged to neutral current rate}

In Sec. \ref{sec2}, we obtained a range of the ratio of charged to 
neutral current event rate at SNO for each of our several cases, 
based on flux-dependent data from SuperKamiokande alone. 
Having identified in Secs. \ref{sec3} and \ref{sec4} 
the allowed shapes of the various $\nu_e$ survival 
probabilities {\it which are now constrained by five experiments}, 
we may narrow these ranges, 
using expressions developed in Sec. \ref{sec2}.  We will approximate the
energy-averaged $\nu_e$ survival probability 
$\langle P^{\oplus}_{ee} \rangle$ as
the $P^{\oplus}_{ee}(E)$ evaluated at an energy of $10 \text{MeV}$ for our
various 
cases.  
The ratios $r_d$ are shown in Table \ref{expectations}.  

Considerable overlapping between the ranges for Cases B1, B2 and C means
that $r_d$ is perhaps not the best experimental observable for their 
disentanglement.  However, by 
measuring $r_d$ alone, these cases are clearly distinguishable from the 
standard $2 \nu$ scenarios, both pure active and pure sterile, and 
from Case A.

\section{Implications of LSND}
\label{lsnd}

The claimed observation of $\overline{\nu}_{\mu} \to \overline{\nu}_e$ 
and $\nu_{\mu}
\to \nu_e$ by LSND suggests small angle mixing between neutrino states 
separated by a
squared mass difference of $\sim 0.1 \to 10\ \text{eV}^2$ \cite{lsnd}.  
This feature
can be easily incorporated into Case A. For Cases B1, B2 and C, and 
adhering to the
standard neutrino mass hierarchy, the required oscillation length suggests 
indirect
$\nu_{\mu} \leftrightarrow \nu_e$ oscillations through a sufficiently heavy
$\nu_{\tau}$ and/or $\nu'_{\tau}$. However, these are quite suppressed if 
consistency
with all other neutrino experiments to date is sought.  Reconciliation with 
the LSND
result occurs at about the $3\sigma$ level in a minute region of parameter 
space (see
Refs.\ \cite{liu} and \cite{grimus} for relevant discussions). 

For greater consistency with LSND, the standard mass hierarchy 
$m_{1\pm} < m_{2\pm} <
m_{3\pm}$ must be altered by interchanging the second and third generation 
neutrinos
(inverse mass hierarchy between $\nu_{\mu}$/$\nu'_{\mu}$ and
$\nu_{\tau}$/$\nu'_{\tau}$).  In this scenario, the MSW partners of 
$\nu_e$ are
$\nu_{\tau}$ and/or $\nu'_{\tau}$.  The $\nu_{\mu}$/$\nu'_{\mu}$ pair is 
now placed
at $\sim 0.1 \to 10\ \text{eV}^2$ above the $\nu_e$/$\nu'_e$ pair on the 
squared mass
spectrum. With an appropriate mixing angle, this will lead to direct 
$\nu_{\mu}
\leftrightarrow \nu_e$ oscillations, thereby accounting for the LSND result.

For both of these scenarios, relic neutrino asymmetry generation 
via ordinary--mirror
neutrino oscillations may not be strong enough to ensure consistency with 
Big Bang
Nucleosynthesis \cite{foot}. The reason is that the LSND $\Delta m^2$ 
tends to be
smaller than the $\Delta m^2$ values favoured by this mechanism 
(see the third paper
in Ref.\cite{bbn}). However, a detailed calculation would need to 
be performed to be
sure of this. If this mechanism fails, then consistency with 
Big Bang Nucleosynthesis
can be obtained by postulating that a sufficiently large neutrino asymmetry
($\stackrel{>}{\sim} 10^{-5}\ \text{eV}^2$) is created at a high 
temperature scale by
some physical mechanism unassociated with ordinary--mirror neutrino 
oscillations
(Ref.\cite{prl} discusses this type of scenario in more detail).

\section{Summary and Conclusions}
\label{sec6}

In order to account for the significant depletion of solar neutrinos 
measured by five
experiments to date, we have invoked MSW-enhanced intergenerational 
mixing in
addition to maximal vacuum $\nu_e \leftrightarrow \nu'_e$ oscillations 
prescribed by
the EPM.  Approximate analytical expressions for the $\nu_e$ survival 
probabilities
as functions of neutrino energy have been obtained for several possible 
neutrino mass
hierarchies assuming small vacuum intergenerational mixing.  These 
expressions were
then compared with well-established $2 \nu$ solutions to identify the 
approximate
regions of parameter space that can simultaneously explain the apparent 
solar and
atmospheric neutrino anomalies.  (Note that only those parameter space 
regions which
feature small matter-induced mixing were examined in depth. 
The approximate allowed
regions plotted in Fig.\ \ref{myfit} assume this restriction. 
Some indication of
the effect of non-negligible matter-induced mixing is discussed 
in Sec.\ \ref{sec3}.) 
The $\nu_e$ survival probabilities for the Cases A,
B1, B2 and C considered herein exhibit considerable differences from 
each other and
from the standard $\nu_e \leftrightarrow \nu_{\mu, \, \tau}$ and $\nu_e
\leftrightarrow \nu_s$ small and large mixing cases. 

These new, distinguishing features are, in principle, observable by future 
solar
neutrino experiments, and have been briefly discussed. The results are 
summarised in
Table V. By utilising Boron neutrino spectral distortion, the Beryllium 
neutrino flux
and the ratio of charged to neutral current event rates, all the EPM 
possibilties can
be distinguished from the standard $2\nu$ MSW solutions, and, with the 
exception of Case C, from models incorporating only one sterile neutrino.
The four cases 
within the
EPM also yield different outcomes, except for Cases B1 and B2 which exhibit an
overlap for these solar neutrino observables. Fortunately, Cases B1 and B2 
can be
differentiated through atmospheric neutrino data. The large 
$\Delta m^2_{1+1-}$ that
defines Case B1 leads to significant $\nu_e \leftrightarrow \nu'_e$
atmospheric neutrino
oscillations. This case is actually already disfavoured by the recent 
atmospheric
neutrino data from SuperKamiokande (see Ref.\cite{bunn} for detailed 
discussions).
Note that Case C within in the EPM is very similar to the scenario analysed in
Ref.\cite{liu}.

We should also note that the implications of MSW solutions within the 
EPM for the
day-night effect have yet to be examined.  In addition, one relevant 
region of
parameter space has not been explored in detail in this paper: 
$\Delta m^2_{1+1-}$
values in the approximate range $10^{-6} \to 10^{-4}\ eV^2$ which are 
intermediate
between Cases B1 and B2. In this regime, the mass-squared difference between
$\nu_{1+}$ and $\nu_{1-}$ is comparable to the MSW intergenerational 
mass difference,
and our approximation scheme is no longer reliable. This region is perhaps best
explored numerically, a task beyond the scope of this paper.

The Exact Parity Model is, in part, an explicit theory of light, 
effectively sterile,
neutrinos. Its characteristic ordinary--mirror neutrino maximal mixing
feature
receives strong experimental support from the atmospheric neutrino data. 
Various
possibilities for solving the solar neutrino problem by either averaged 
vacuum $\nu_e
\leftrightarrow \nu'_e$ oscillations alone (Case A), or several amalgams of
MSW-enhanced and vacuum oscillations (Cases B1, B2 and C), exist within the 
EPM.
Future solar neutrino experiments should narrow the possibilities considerably.

\acknowledgments{This work was supported in part by the Australian Research 
Council
and in part by the Commonwealth of Australia's postgraduate award scheme.
YYYW would like to thank John A.\ L.\ McIntosh for asssistance with
preparation of the
Figures.}

\begin{table}
\caption{Solar neutrino measurements and theoretical expectations within 
the standard solar model of Bahcall and Pinsonneault (SSM-BP) (1995)
\protect\cite{bp95}.
Capture rates for Homestake \protect\cite{homestake}, GALLEX
\protect\cite{gallex} and SAGE \protect\cite{sage} are given in SNU, where 
$1\, \text{SNU} = 10^{-36}\text{capture per atom per second}$.  
For Kamiokande \protect\cite{kamioka} and SuperKamiokande
\protect\cite{superk}, the measured neutrino flux 
is given in $10^6\text{cm}^{-2}\text{s}^{-1}$.  The associated statistical and 
systematic errors ($1 \sigma$) are quoted for each experiment.}
\label{expresults}
\begin{tabular}{lll}
Experiment & Measurement & SSM-BP  \\
\tableline
Homestake & $2.56 \pm 0.16 \pm 0.14$ & $9.3^{+1.2}_{-1.4}$ \\
GALLEX & $69.7 \pm 6.7^{+3.9}_{-4.5}$ & $137^{+8}_{-7}$  \\
SAGE & $72^{+12+5}_{-10-7}$ & $137^{+8}_{-7}$  \\
Kamiokande &$2.80 \pm 0.19 \pm 0.33$ & $6.62^{+0.93}_{-1.12}$ \\
SuperKamiokande & $2.51^{+0.14}_{-0.13} \pm 0.18$ & $6.62^{+0.93}_{-1.12}$
\end{tabular}
\vspace{15mm}

\caption{The mixing angle $\eta$, the effective density $A_{eff}$ and 
the off-diagonal 
matter-induced mixing term $A_{ind}$ evaluated for various combinations
of $\nu_{1h}$ and $\nu_{2 \ell}$ for Case B1.}
\label{bonetable} 
\begin{tabular}{cccc}
$\nu_{1h} \leftrightarrow \nu_{2 \ell}$ & $\sin \eta$& $A_{eff}$ & $A_{ind}$ \\
\tableline
$\nu_{1-} \leftrightarrow \nu_{2-}$ & $\sin \phi$ & $\frac{A_{CC}}{2}$ & $0$ \\
$\nu_{1-} \leftrightarrow \nu_{2+}$ 
		& $\frac{\sin \theta}{\sqrt{\cos^2 \phi + \sin^2 \theta}}$ 
		& $\frac{A_{CC}}{2} (\cos^2 \phi + \sin^2 \theta)  
		+ A_{NC} \sin 2 \eta \sin (\theta - \phi)$
		& $A_{NC} \cos 2 \eta \sin (\theta - \phi)$ \\
$\nu_{1+} \leftrightarrow \nu_{2+}$&$\sin \theta$ & $\frac{A_{CC}}{2}$ & $0$\\
$\nu_{1+} \leftrightarrow \nu_{2-}$ 
		& $\frac{\sin \phi}{\sqrt{\cos^2 \theta + \sin^2 \phi}}$
		& $\frac{A_{CC}}{2} (\cos^2 \theta + \sin^2 \phi)
		- A_{NC} \sin 2 \eta \sin (\theta - \phi)$ 
		& $- A_{NC} \cos 2 \eta \sin (\theta - \phi)$\\
\end{tabular}
\vspace{15mm}

\caption{The mixing angles $\psi$ and $\eta$, the effective density $A_{eff}$
and the off-diagonal matter-induced mixing term $A_{ind}$
evaluated for various combinations of $\nu_{1 \ell}, \, \nu_{1h}$ and 
$\nu_{2 \ell}$ for Case B2.}
\label{btwotable}
\begin{tabular}{ccccc}
$\begin{array}{c}
	\nu_{1 \ell} \\
	 \nu_{1h} \\
	\end{array} \leftrightarrow \nu_{2 \ell}$
	& $\sin \psi$ & $\sin \eta$ & $A_{eff}$ & $A_{ind}$ \\
\tableline
$\begin{array}{c}
	\nu_{1+} \\
	 \nu_{1-} \\
	\end{array}
	\leftrightarrow \nu_{2-}$
	& $\frac{\cos \phi}{\sqrt{\cos^2 \theta + \cos^2 \phi}}$
	& $\frac{ \sin \phi}{\sqrt{1 + \cos^2 \theta}}$
	& $\begin{array}{l}
		\frac{A_{CC}}{2}(1 + \cos^2 \theta) \\
	+ \frac{A_{NC}}{2}[\cos 2 \eta \sin 2 \psi \cos (\theta - \phi) \\
	- 2 \sin 2 \eta \cos \psi \sin (\theta - \phi)] \\
	\end{array}$ 
	& $ \begin{array}{c}
	- A_{NC} [\cos 2 \eta \cos \psi \sin(\theta - \phi) \\
	+ \frac{1}{2} \sin 2 \eta \sin 2 \psi \cos (\theta -\phi)]\\
	\end{array}$ \\
\tableline
$\begin{array}{c}
	\nu_{1-} \\
	 \nu_{1+} \\
	\end{array}
	\leftrightarrow \nu_{2-}$
	& $\frac{\cos \theta}{\sqrt{\cos^2 \theta + \cos^2 \phi}}$
	& $\frac{ \sin \phi}{\sqrt{1 + \cos^2 \theta}}$
	& $\begin{array}{l}
		\frac{A_{CC}}{2}(1 + \cos^2 \theta) \\
	+ \frac{A_{NC}}{2}[\cos 2 \eta \sin 2 \psi \cos (\theta - \phi) \\
	- 2 \sin 2 \eta \sin \psi \sin (\theta - \phi)] \\
	\end{array}$ 
	& $ \begin{array}{c}
	- A_{NC} [\cos 2 \eta \sin \psi \sin(\theta - \phi) \\
	+ \frac{1}{2} \sin 2 \eta \sin 2 \psi \cos (\theta -\phi)]\\
	\end{array}$ \\
\tableline
$\begin{array}{c}
	\nu_{1-} \\
	 \nu_{1+} \\
	\end{array}
	\leftrightarrow \nu_{2+}$
	& $\frac{\cos \theta}{\sqrt{\cos^2 \theta + \cos^2 \phi}}$
	& $\frac{ \sin \theta}{\sqrt{1 + \cos^2 \phi}}$
	& $\begin{array}{l}
		\frac{A_{CC}}{2}(1 + \cos^2 \phi) \\
	+ \frac{A_{NC}}{2}[\cos 2 \eta \sin 2 \psi \cos (\theta - \phi) \\
	+ 2 \sin 2 \eta \cos \psi \sin (\theta - \phi)] \\
	\end{array}$ 
	& $ \begin{array}{c}
	A_{NC} [\cos 2 \eta \cos \psi \sin(\theta - \phi) \\
	- \frac{1}{2} \sin 2 \eta \sin 2 \psi \cos (\theta -\phi)]\\
	\end{array}$ \\
\tableline
$\begin{array}{c}
	\nu_{1+} \\
	 \nu_{1-} \\
	\end{array}
	\leftrightarrow \nu_{2+}$
	& $\frac{\cos \phi}{\sqrt{\cos^2 \theta + \cos^2 \phi}}$
	& $\frac{ \sin \theta}{\sqrt{1 + \cos^2 \phi}}$
	& $\begin{array}{l}
		\frac{A_{CC}}{2}(1 + \cos^2 \phi) \\
	+ \frac{A_{NC}}{2}[\cos 2 \eta \sin 2 \psi \cos (\theta - \phi) \\
	+ 2 \sin 2 \eta \sin \psi \sin (\theta - \phi)] \\
	\end{array}$ 
	& $ \begin{array}{c}
	A_{NC} [\cos 2 \eta \sin \psi \sin(\theta - \phi) \\
	- \frac{1}{2} \sin 2 \eta \sin 2 \psi \cos (\theta -\phi)]\\
	\end{array}$ \\
\end{tabular}
\vspace{15mm}

\caption{The allowed intergenerational oscillation parameters $\Delta m^2$
and $\sin^2 2 \eta$ for Cases B1,
B2 and C.  The best fit oscillation parameters in the small mixing angle 
(SMA) and the large mixing angle (LMA)
solutions for standard $2\nu$ scenarios \protect\cite{hl,earth} are
also included.}
\label{summary}
\begin{tabular}{ccc}
Scheme & $\Delta m^2/eV^2$ & $\sin^2 2 \eta$ \\
\tableline
Standard $\nu_e \leftrightarrow \nu_{\mu, \, \tau}$ SMA
& $5 \times 10^{-6}$ & $8 \times 10^{-3}$ \\
Standard $\nu_e \leftrightarrow \nu_{\mu, \, \tau}$ LMA
& $1.6 \times 10^{-5}$ & $0.63$ \\
Standard $\nu_e \leftrightarrow \nu_s$ SMA
& $4 \times 10^{-6}$ & $10^{-2}$ \\
Case B1 & $ 2 \times 10^{-6} \to 5 \times 10^{-5}$ 
& $8 \times 10^{-4} \to 0.1$ \\ 
Case B2 & $ 10^{-5} \to 10^{-4}$ & $10^{-4} \to 4 \times 10^{-3}$ \\
Case C & $4 \times 10^{-6} \to 5 \times 10^{-6}$
& $8 \times 10^{-3} \to 10^{-2}$ \\
\end{tabular} 

\vspace{15mm}
\caption{Predictions for future experiments.  The amount of Boron spectral
distortion associated with each case is qualitatively compared with the 
standard $2 \nu$ predictions (maximal for SMA and minimal for LMA).  
The approximate Beryllium fluxes to be measured by BOREXINO prescribed
by our various cases relative to the no-oscillation flux are also compared.
The last column shows the predicted ranges of the ratio of charged to 
neutral current event rate relative to the no-oscillation rate at SNO.}
\label{expectations}
\begin{tabular}{cccc}
Scheme & Boron spectral distortion & Be flux 
($\frac{\text{predicted}}{\text{SSM}}$)
& $r_d$ for SNO \\
\tableline
Standard $\nu_e \leftrightarrow \nu_{\mu, \, \tau}$ SMA
& maximal & $\sim 0$ & $0.25 \to 0.4$ \\
Standard $\nu_e \leftrightarrow \nu_{\mu, \, \tau}$ LMA 
& minimal & $\sim 0.4$ & $\sim 0.2$ \\
Standard $\nu_e \leftrightarrow \nu_s$ SMA 
& maximal & $\sim 0$ & $1$ \\
Case A & none & $\frac{1}{2} $ & $1$ \\
Case B1 & intermediate & $\frac{1}{4} \to \frac{1}{2}$
& $0.5 \to 0.75$ \\
Case B2 & intermediate & $\sim \frac{1}{2}$ & $0.4 \to 0.7$ \\
Case C & maximal & $\sim 0$ & $0.45 \to 0.6$ \\
\end{tabular}

\end{table}

\begin{figure}
\caption{Level-crossing diagram for Case B1.  The labels $\nu^m_{i}$, where
$i = 1 \ell, \, 1h, \, 2 \ell, \, 2h$, denote the instantaneous mass 
eigenstates.  The letter $R$ labels the intergenerational MSW resonance.}
\label{lsl}
\vspace{5mm}

\caption{The $\nu_e$ survival probability at the Earth for Case B1 
with $\Delta m^2_{21} = 5 \times 10^{-6} eV^2$ and
$\sin^2 2 \eta = 8 \times 10^{-3}$ (solid line) for $\nu_e$ produced at the
centre of the sun.  
The dashed and dotted lines represent, 
respectively, the survival probabilities for the standard 
$\nu_e \leftrightarrow \nu_{\mu,\,\tau}$ and $\nu_e \leftrightarrow \nu_s$
scenarios evaluated for the same oscillation parameters.}
\label{lslprob}
\vspace{5mm}

\caption{Level-crossing diagram for Case B2.  The labels $\nu^m_{i}$, where
$i = 1 \ell, \, 1h, \, 2 \ell, \, 2h$, denote the instantaneous mass 
eigenstates.  Two resonances are identified and labelled as $R_L$ and $R_H$,
where $R_H$ is the resonance of interest.}
\label{lss}
\vspace{5mm}

\caption{The $\nu_e$ survival probability at the Earth for Case B2 
with $\Delta m^2_{21} = 5 \times 10^{-6} eV^2$ and 
$\sin^2 2 \eta = 8 \times 10^{-3}$ (solid line) for $\nu_e$ produced at the 
centre of the sun.  
The dashed and dotted lines represent, 
respectively, the survival probabilities for the standard 
$\nu_e \leftrightarrow \nu_{\mu,\,\tau}$ and $\nu_e \leftrightarrow \nu_s$
scenarios evaluated for the same oscillation parameters.}
\label{lssprob}
\vspace{5mm}

\caption{The $\nu_e$ survival probability at the Earth for Case C
with $\Delta m^2_{21} = 5 \times 10^{-6} eV^2$ and
$\sin^2 2 \eta = 8 \times 10^{-3}$ (solid line) for $\nu_e$ produced at the
centre of the sun.  
The dashed and dotted lines represent, 
respectively, the survival probabilities for the standard 
$\nu_e \leftrightarrow \nu_{\mu,\,\tau}$ and $\nu_e \leftrightarrow \nu_s$
scenarios evaluated for the same oscillation parameters.}
\label{novac}
\vspace{5mm}

\caption{The approximate allowed regions for Cases B1 (area enclosed by
the dot-dash line) and B2 (area enclosed by the dashed line).  These 
correspond to the regions in which the $2\sigma$ bands 
(including theoretical uncertainties) of all five solar neutrino experiments 
(see Table \ref{expresults}) overlap, respectively for Cases B1 and B2.
Note that these regions will differ slightly from those generated from a 
$\chi^2$-analysis.  The allowed region for the standard $2\nu$ case
at $95 \%$ C.L. (dotted line) is shown here for the purpose of comparison 
\protect\cite{hl}.  The solution to Case C is similar to this.}
\label{myfit}
\vspace{5mm}

\caption{The $\nu_e$ survival probability at the Earth for Case B1 
evaluated for various $\Delta m^2_{21}$ and $\sin^2 2 \eta$ shown on the 
graphs (solid line) for $\nu_e$ produced at the centre of the sun.  
These parameters lie within the allowed 
region for Case B1 but are not necessarily the best fit parameters.
For the purpose of comparison, the survival probabilities for the standard 
$\nu_e \leftrightarrow \nu_{\mu,\,\tau}$ small angle solution (dashed line) 
and large angle solution (dotted line)  
(see Table \ref{summary} for the best fit parameters) are also shown.}
\label{lslfit}
\vspace{5mm}

\caption{The $\nu_e$ survival probability at the Earth for Case B2 
evaluated for various $\Delta m^2_{21}$ and $\sin^2 2 \eta$ shown on the
graphs (solid line) for $\nu_e$ produced at the centre of the sun.  
These parameters lie within the allowed 
region for Case B2 but are not necessarily the best fit parameters.
For the purpose of comparison, the survival probabilities for the standard 
$\nu_e \leftrightarrow \nu_{\mu,\,\tau}$ small angle solution (dashed line) 
and large angle solution (dotted line) (see Table \ref{summary} for the best
fit parameters) are also shown.}
\label{lssfit}

\newpage
\epsfig{file=lsl.epsi,width=15cm}
\vspace{5mm}
\begin{center}
FIG. \ref{lsl}
\end{center}

\newpage
\epsfig{file=lslprob.epsi,width=15cm}
\vspace{5mm}
\begin{center}
FIG. \ref{lslprob}
\end{center}

\newpage
\epsfig{file=lss.epsi,width=15cm}
\vspace{5mm}
\begin{center}
FIG. \ref{lss}
\end{center}

\newpage
\epsfig{file=lssprob.epsi,width=15cm}
\vspace{5mm}
\begin{center}
FIG. \ref{lssprob}
\end{center}

\newpage
\epsfig{file=novac.epsi,width=15cm}
\vspace{5mm}
\begin{center}
FIG. \ref{novac}
\end{center}

\newpage
\epsfig{file=myfit.epsi,width=15cm}
\vspace{5mm}
\begin{center}
FIG. \ref{myfit}
\end{center}

\newpage
\epsfig{file=lslfit.epsi,width=15cm}
\vspace{5mm}
\begin{center}
FIG. \ref{lslfit}
\end{center}

\newpage
\epsfig{file=lssfit.epsi,width=15cm}
\vspace{5mm}
\begin{center}
FIG. \ref{lssfit}
\end{center}

\end{figure}

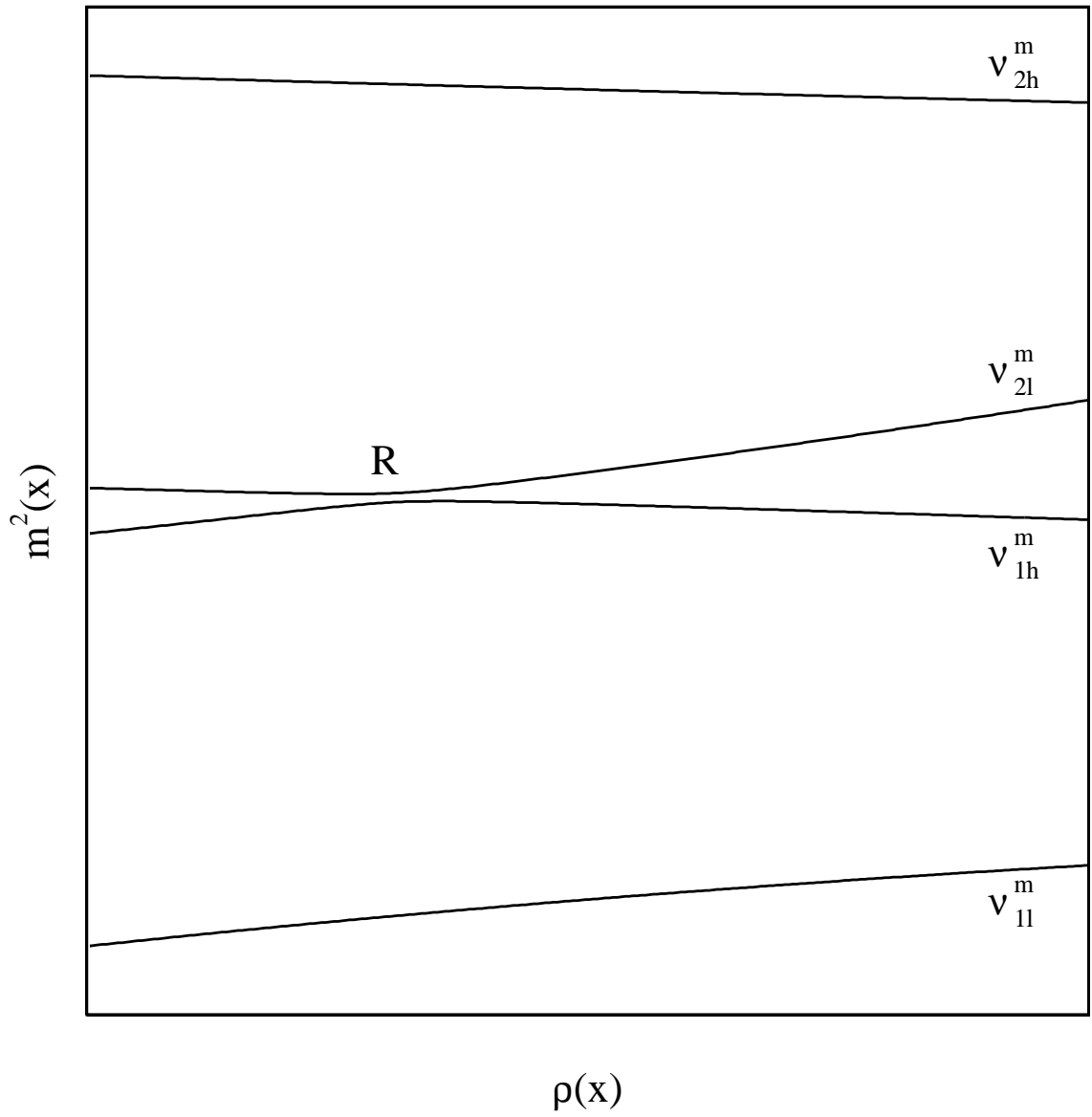
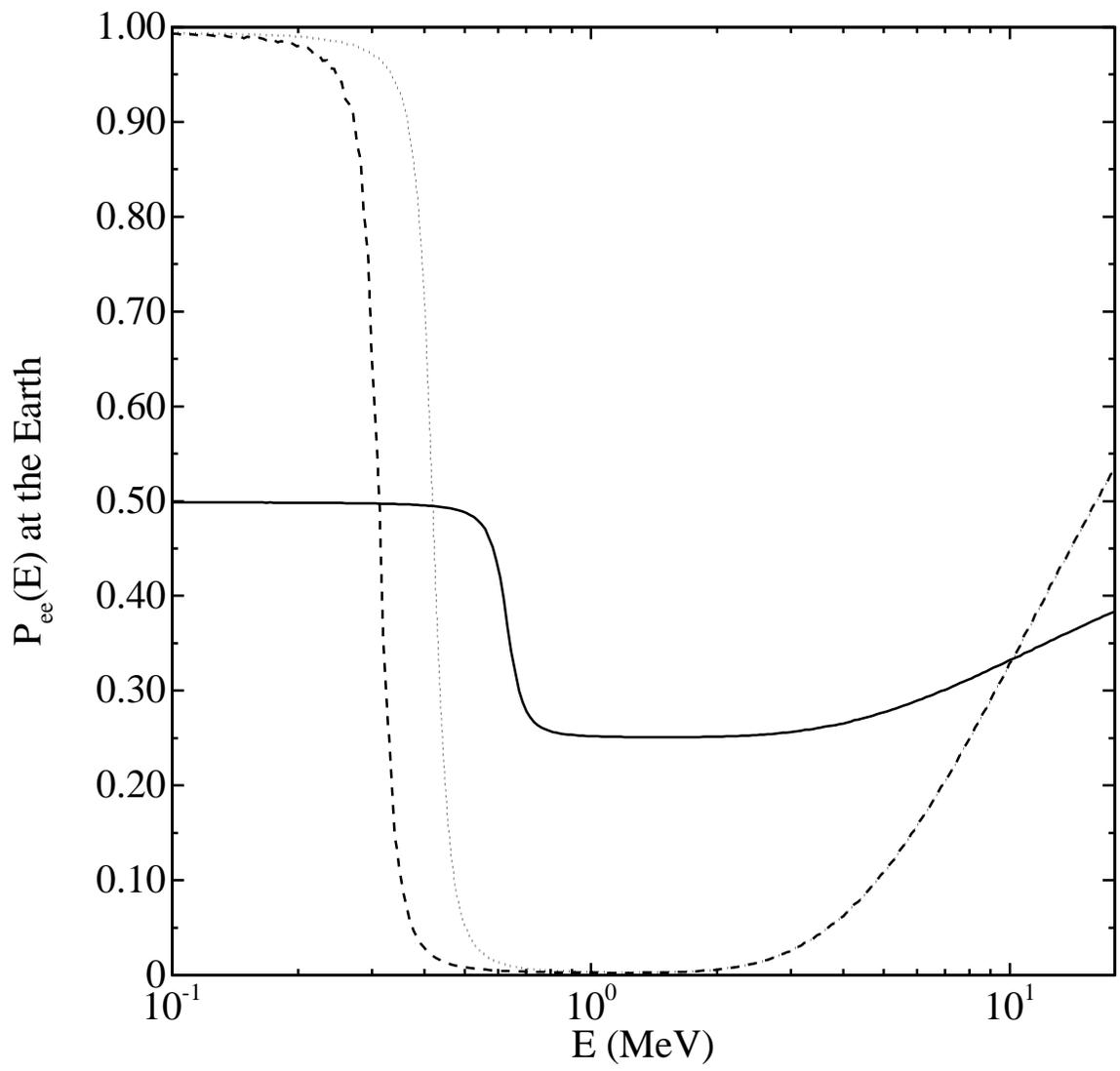
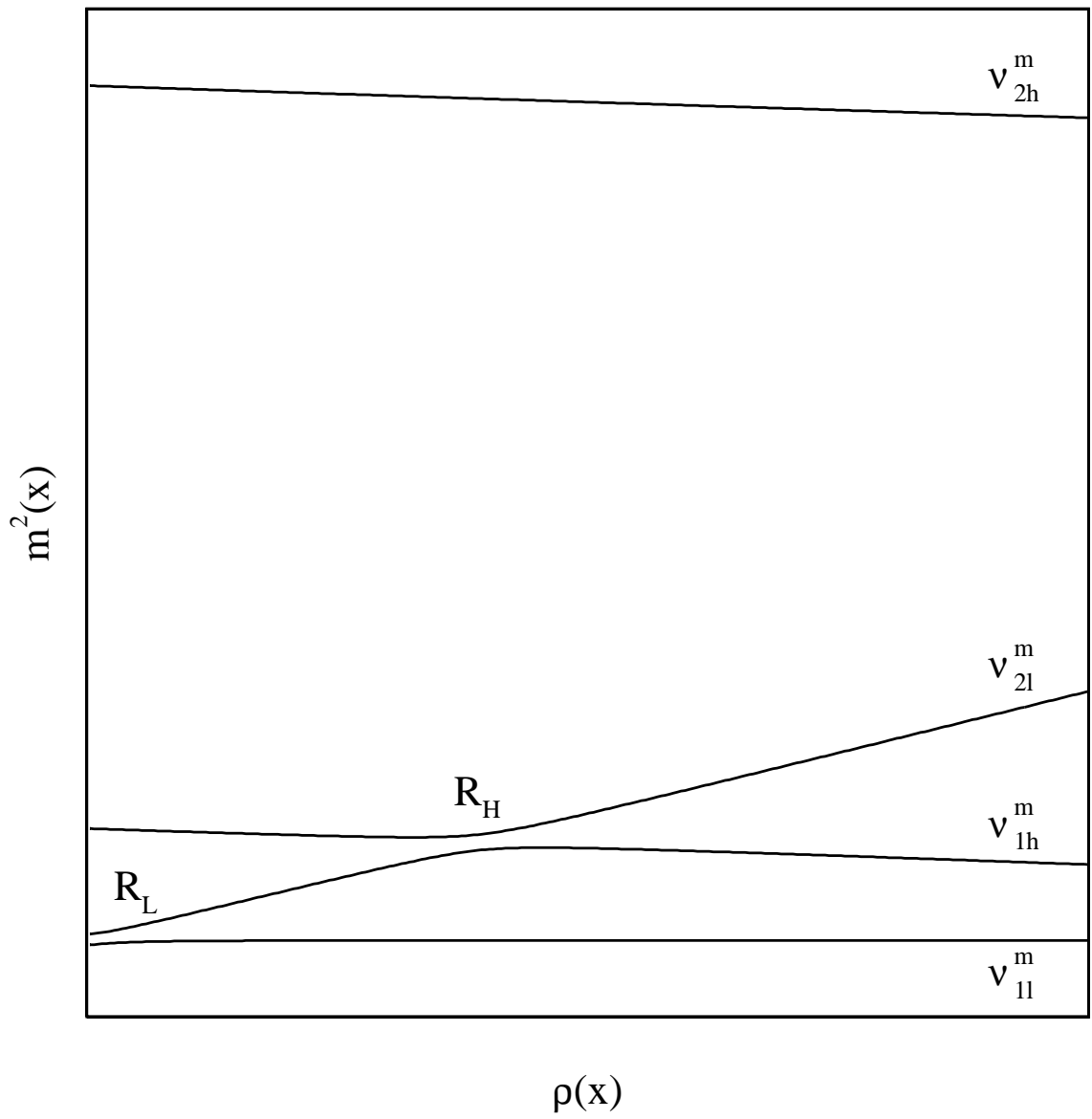
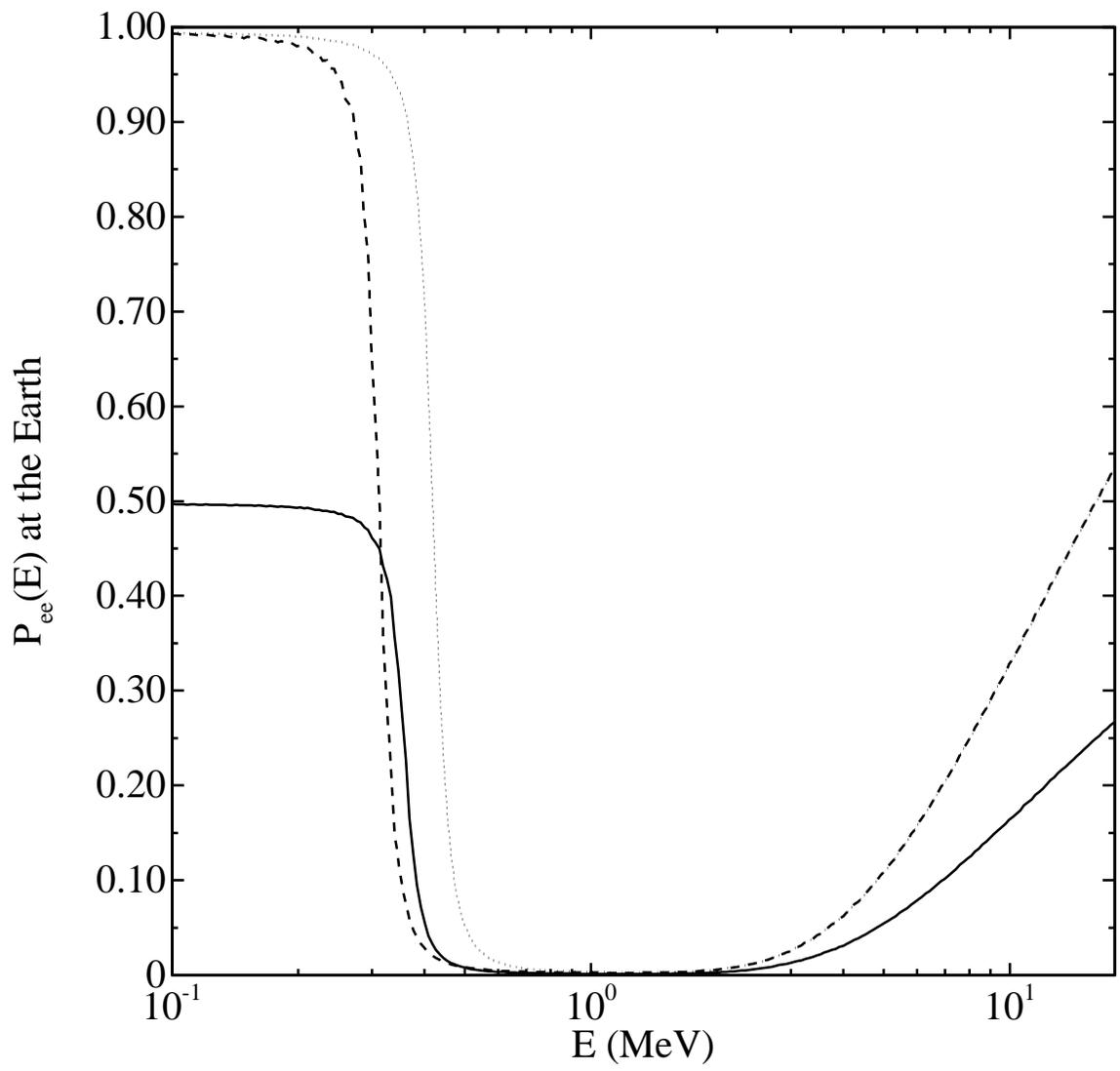
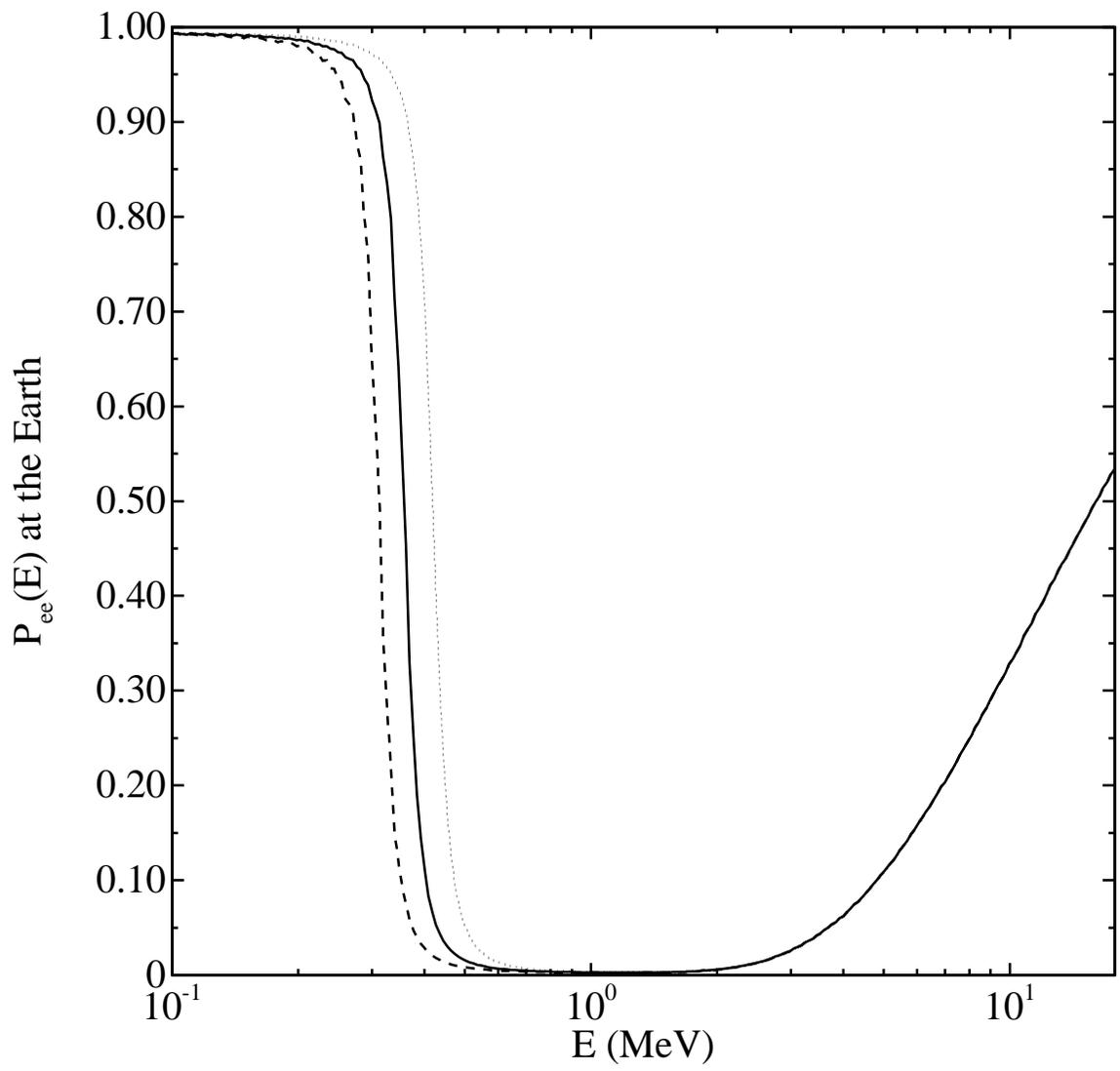
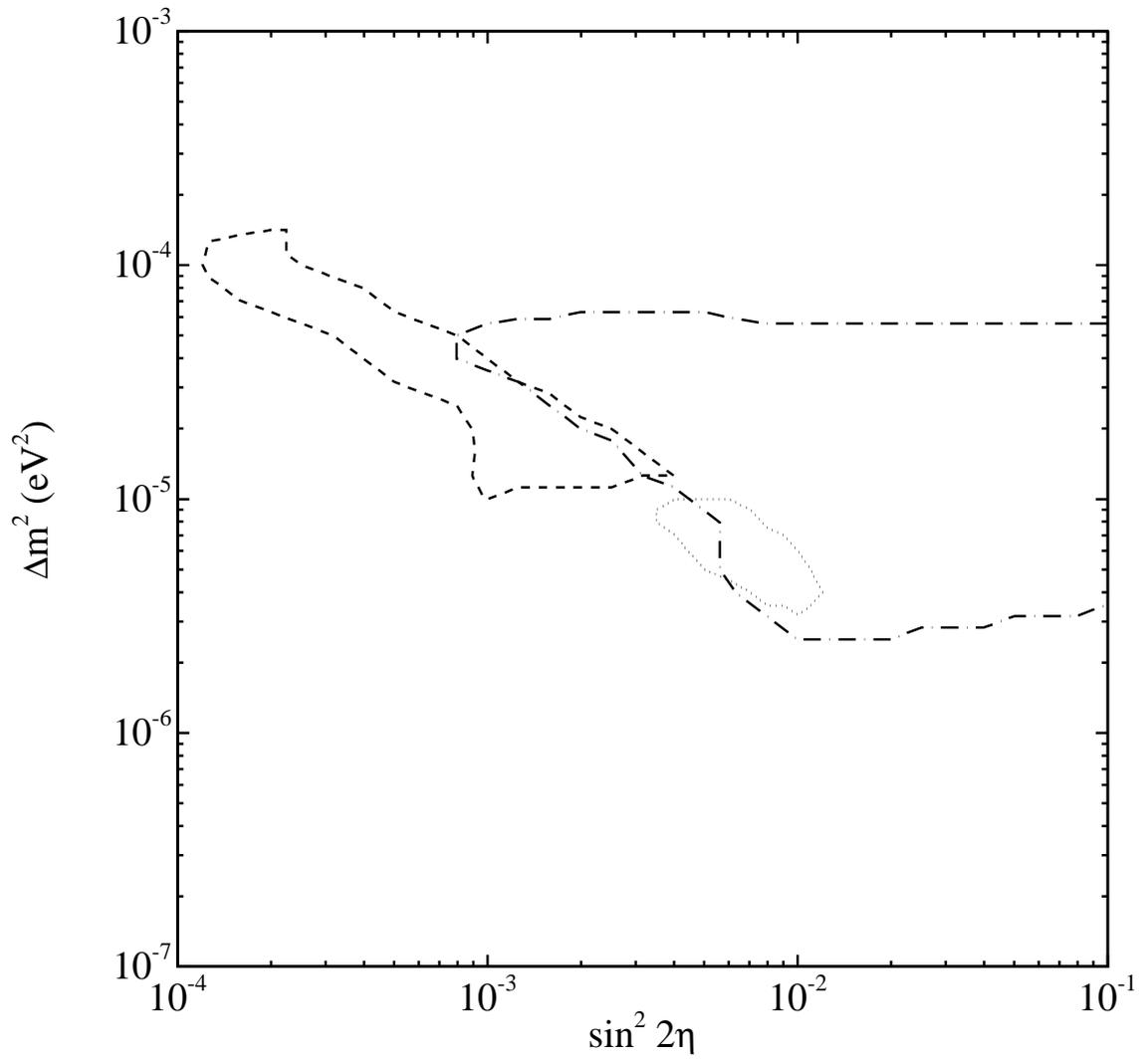
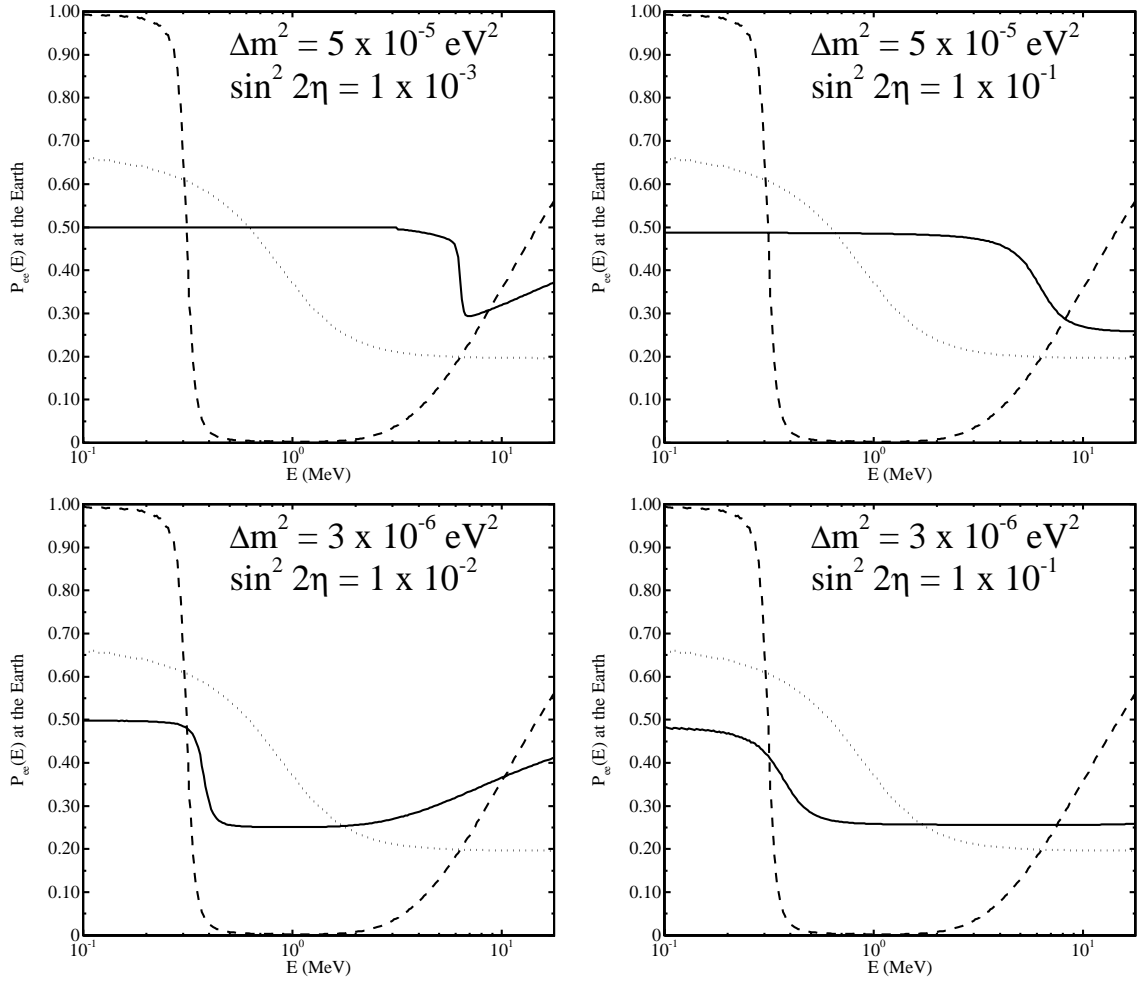
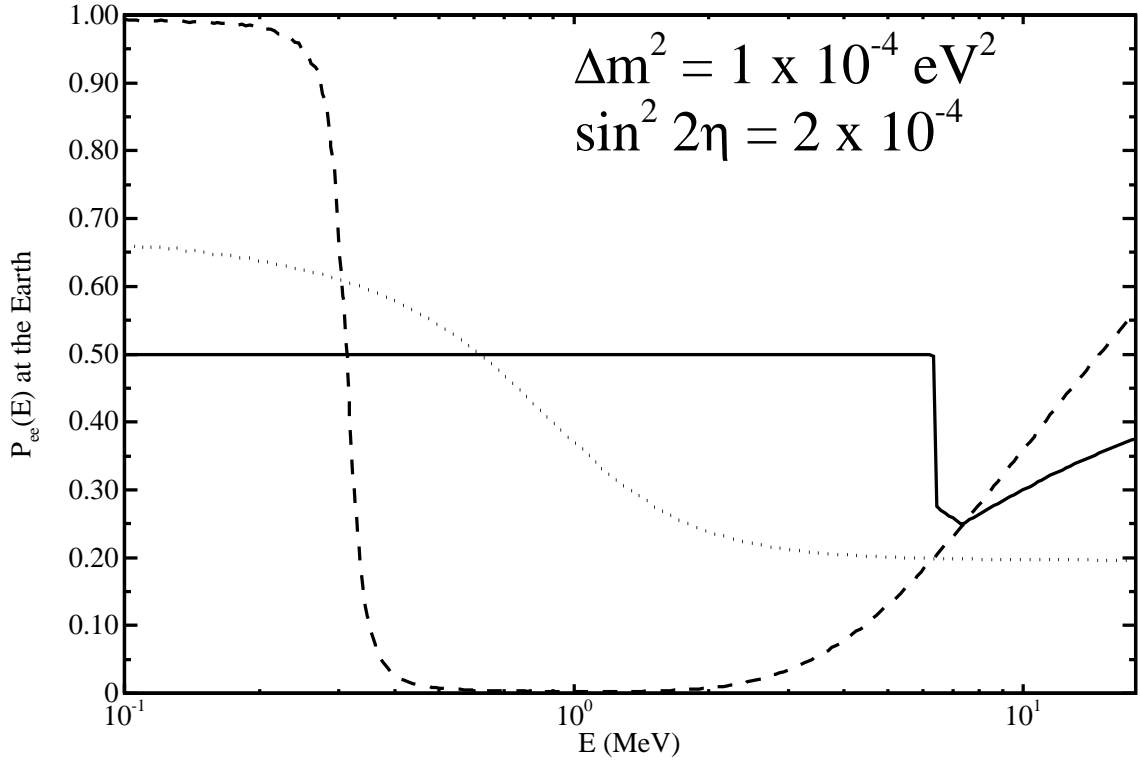
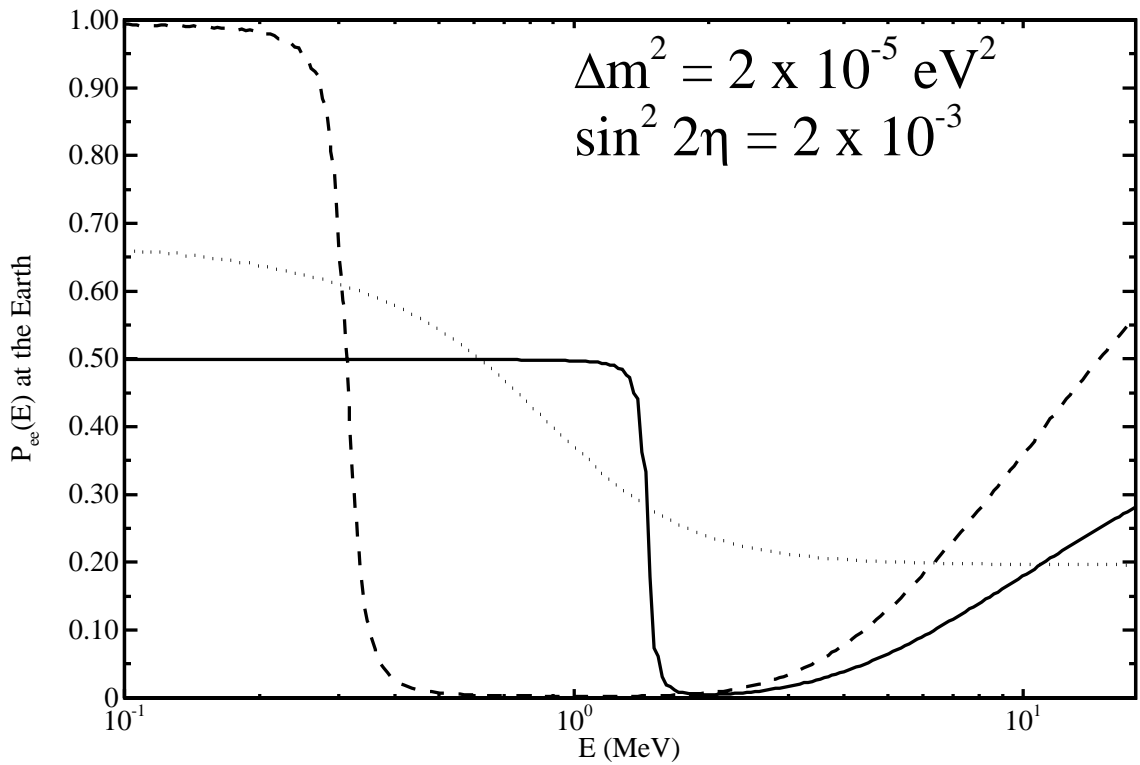
\end{document}